\documentclass[amsmath,amssymb,superscriptaddress,longbibliography,nobalancelastpage,aps,twocolumn,showpacs]{revtex4-1}

\usepackage{array,multirow,graphicx}
\usepackage{braket}
\usepackage{epstopdf}
\usepackage{graphicx}
\usepackage{siunitx}
\usepackage{nicefrac}
\usepackage[normalem]{ulem}
\usepackage{varioref}
\usepackage{xcolor}
\usepackage{xfrac}
\usepackage{xr-hyper}
\usepackage{hyperref}

\definecolor{bl}{rgb}{0.0,0.2,0.6}

\hypersetup{colorlinks,linkcolor=blue,urlcolor=blue,citecolor=blue}

\newcommand{\CRO}{Ca$_2$RuO$_4$}
\newcommand{\SRO}{Sr$_2$RuO$_4$}
\newcommand{\SIO}{Sr$_2$IrO$_4$}

\newcommand{\EF}{$E_\mathrm{F}$}

\newcommand{\dxy}{$d_{xy}$}
\newcommand{\dxz}{$d_{xz}$}
\newcommand{\dyz}{$d_{yz}$}

\begin{document}
	\author{M.~Horio}
	\email{mhorio@issp.u-tokyo.ac.jp}
	\affiliation{Institute for Solid State Physics, The University of Tokyo, Kashiwa, Chiba 277-8581, Japan}
	\affiliation{Physik-Institut, Universit\"{a}t Z\"{u}rich, Winterthurerstrasse 190, CH-8057 Z\"{u}rich, Switzerland}
	
	\author{F.~Forte}
	\email{filomena.forte@spin.cnr.it}
	\affiliation{CNR-SPIN, I-84084 Fisciano, Salerno, Italy}
	\affiliation{Dipartimento di Fisica "E.R.~Caianiello", Universit\`{a} di Salerno, I-84084 Fisciano, Salerno, Italy}
	
	\author{D.~Sutter}
	\affiliation{Physik-Institut, Universit\"{a}t Z\"{u}rich, Winterthurerstrasse 190, CH-8057 Z\"{u}rich, Switzerland}
	
	\author{M.~Kim}
	\affiliation{Coll\`ege de France, F-75231 Paris Cedex 05, France}
	\affiliation{Centre de Physique Th\'{e}orique, Ecole Polytechnique, CNRS, Universit\'{e} Paris-Saclay, F-91128 Palaiseau, France}
	\affiliation{Korea Institute for Advanced Study, Seoul 02455, South Korea}
	
	\author{C.~G.~Fatuzzo}
	\affiliation{Physik-Institut, Universit\"{a}t Z\"{u}rich, Winterthurerstrasse 190, CH-8057 Z\"{u}rich, Switzerland}
	
	\author{C.~E.~Matt}
	\affiliation{Physik-Institut, Universit\"{a}t Z\"{u}rich, Winterthurerstrasse 190, CH-8057 Z\"{u}rich, Switzerland}
	
	\author{S.~Moser}
	\affiliation{Advanced Light Source (ALS), Berkeley, California 94720, USA}
	\affiliation{Physikalisches Institut and Würzburg-Dresden Cluster of Excellence ct.qmat, Universität Würzburg, Würzburg 97074, Germany}
	
	\author{T.~Wada}
	\affiliation{Institute for Solid State Physics, The University of Tokyo, Kashiwa, Chiba 277-8581, Japan}
	
	\author{V.~Granata}
	\affiliation{Dipartimento di Fisica "E.R.~Caianiello", Universit\`{a} di Salerno, I-84084 Fisciano, Salerno, Italy}
	
	\author{R.~Fittipaldi}
	\affiliation{CNR-SPIN, I-84084 Fisciano, Salerno, Italy}
	\affiliation{Dipartimento di Fisica "E.R.~Caianiello", Universit\`{a} di Salerno, I-84084 Fisciano, Salerno, Italy}
	
	\author{Y.~Sassa}
	\affiliation{Department of Physics, Chalmers University of Technology, SE-412 96 G\"{o}teborg, Sweden}
	
	\author{G.~Gatti}
	\affiliation{Institute of Physics, \'{E}cole Polytechnique Fed\'{e}rale de Lausanne (EPFL), CH-1015 Lausanne, Switzerland}
	\affiliation{Department of Quantum Matter Physics, University of Geneva, 24 Quai Ernest-Ansermet, CH-1211 Geneva 4, Switzerland}
	
	
	\author{H.~M.~R\o nnow}
	\affiliation{Institute of Physics, \'{E}cole Polytechnique Fed\'{e}rale de Lausanne (EPFL), CH-1015 Lausanne, Switzerland}
	
	\author{M.~Hoesch}
	\affiliation{Diamond Light Source, Harwell Campus, Didcot, OX11 0DE, United Kingdom}
	\affiliation{Photon Science, Deutsches Elektronen-Synchrotron (DESY), Notekestrasse 85, 22607 Hamburg, Germany}
	
	\author{T.~K.~Kim}
	\affiliation{Diamond Light Source, Harwell Campus, Didcot, OX11 0DE, United Kingdom}
	
	\author{C.~Jozwiak}
	\affiliation{Advanced Light Source (ALS), Berkeley, California 94720, USA}
	
	\author{A.~Bostwick}
	\affiliation{Advanced Light Source (ALS), Berkeley, California 94720, USA}
	
	\author{Eli~Rotenberg}
	\affiliation{Advanced Light Source (ALS), Berkeley, California 94720, USA}
	
	\author{I.~Matsuda}
	\affiliation{Institute for Solid State Physics, The University of Tokyo, Kashiwa, Chiba 277-8581, Japan}

	\author{A.~Georges}
	\affiliation{Coll\`ege de France, F-75231 Paris Cedex 05, France}
	\affiliation{Centre de Physique Th\'{e}orique, Ecole Polytechnique, CNRS, Universit\'{e} Paris-Saclay, F-91128 Palaiseau, France}
	\affiliation{Department of Quantum Matter Physics, University of Geneva, 24 Quai Ernest-Ansermet, CH-1211 Geneva 4, Switzerland}
	\affiliation{Center for Computational Quantum Physics, Flatiron Institute, 162 5th Avenue, New York, New York 10010, USA}
	
	\author{G.~Sangiovanni}
	\affiliation{Institut f\"ur Theoretische Physik und Astrophysik and W\"urzburg-Dresden Cluster of Excellence ct.qmat, Universit\"at W\"urzburg, W\"urzburg, Germany}
	
	\author{A.~Vecchione}
	\affiliation{CNR-SPIN, I-84084 Fisciano, Salerno, Italy}
	\affiliation{Dipartimento di Fisica "E.R.~Caianiello", Universit\`{a} di Salerno, I-84084 Fisciano, Salerno, Italy}
	
	\author{M.~Cuoco}
	\affiliation{CNR-SPIN, I-84084 Fisciano, Salerno, Italy}
	\affiliation{Dipartimento di Fisica "E.R.~Caianiello", Universit\`{a} di Salerno, I-84084 Fisciano, Salerno, Italy}
	
	\author{J.~Chang}
	\affiliation{Physik-Institut, Universit\"{a}t Z\"{u}rich, Winterthurerstrasse 190, CH-8057 Z\"{u}rich, Switzerland}

	\title{Orbital-selective 
		metal skin induced by alkali-metal-dosing Mott-insulating \CRO}

	\maketitle

	\textbf{Doped Mott insulators are the starting point for interesting physics such as high temperature superconductivity and quantum spin liquids. For multi-band Mott insulators, orbital selective ground states have been envisioned. However, orbital selective metals and Mott insulators have  been difficult to realize experimentally. Here we demonstrate by photoemission spectroscopy how \CRO, upon alkali-metal surface doping, develops a single-band metal skin. Our dynamical mean field theory calculations reveal that homogeneous electron doping of \CRO\ results in a multi-band metal. All together, our results  provide compelling evidence for an orbital-selective Mott insulator breakdown, which  is unachievable via simple electron doping. Supported by a cluster model and cluster perturbation theory calculations, we demonstrate a novel type of skin metal-insulator transition induced by surface dopants that orbital-selectively hybridize with the bulk Mott state and in turn produce coherent in-gap states.} 


	\vspace{5mm}
	\textbf{Introduction}
	
	Interface metallicity paves the way for two-dimensional fermionic gasses with interesting properties such as superconductivity~\cite{CavigliaNat2008,KimNatPhys2016,LiuScience2021}. Insulators are building blocks for such interfaces. Upon doping (charge transfer), two-dimensional metals can occur at the interface between two insulators like LaAlO$_3$ and SrTiO$_3$~\cite{LiuNatComm2023,ReyrenScience2007} or at the insulator-vacuum interface like the surface of SrTiO$_3$~\cite{SantanderNature2011,MoserPRL2013}. 
	Such metallic states confined at interfaces are broadly dubbed quantum well states. This term is used irrespectively of insulating nature (band or Mott insulators).
	For many spectroscopies, insulator - vacuum interfaces are interesting as they are directly accessible 
	in contrast to buried~\cite{Woerle2017} insulator-insulator quantum wells. Doping of insulator-vacuum interfaces are often achieved by dosing with alkali metal atoms~\cite{KimScience2014,AlidoustNatCommun2014,EknapakulPRB2016,AkikuboTSF2016,YukawaPRB2018}. 
	Electrons from the alkali-metal layer can form a quantum-well state confined by vacuum and the band gap of the substrate. When the substrate band gap is sufficiently large, the metallic state can be strictly confined inside the alkali-metal overlayer~\cite{EknapakulPRB2016}. On the other hand, with significant energy and momentum overlap, the quantum well state will hybridize with the substrate and form a hybrid state~\cite{AlidoustNatCommun2014,AkikuboTSF2016}. In this context, particularly interesting is the case of Mott insulators as a substrate. In contrast to the rigid band gap of band insulators, a Mott gap is maintained by a delicate balance between kinetic energy and electron correlation~\cite{ImadaRMP1998}. This in turn suggests that the interaction between the alkali metal and the electronic states of the Mott insulator could trigger the breakdown of the Mott state at the surface, leaving a hybrid quantum-well state.

	\begin{figure*}[ht!]
		\begin{center}
			\includegraphics[width=0.9\textwidth]{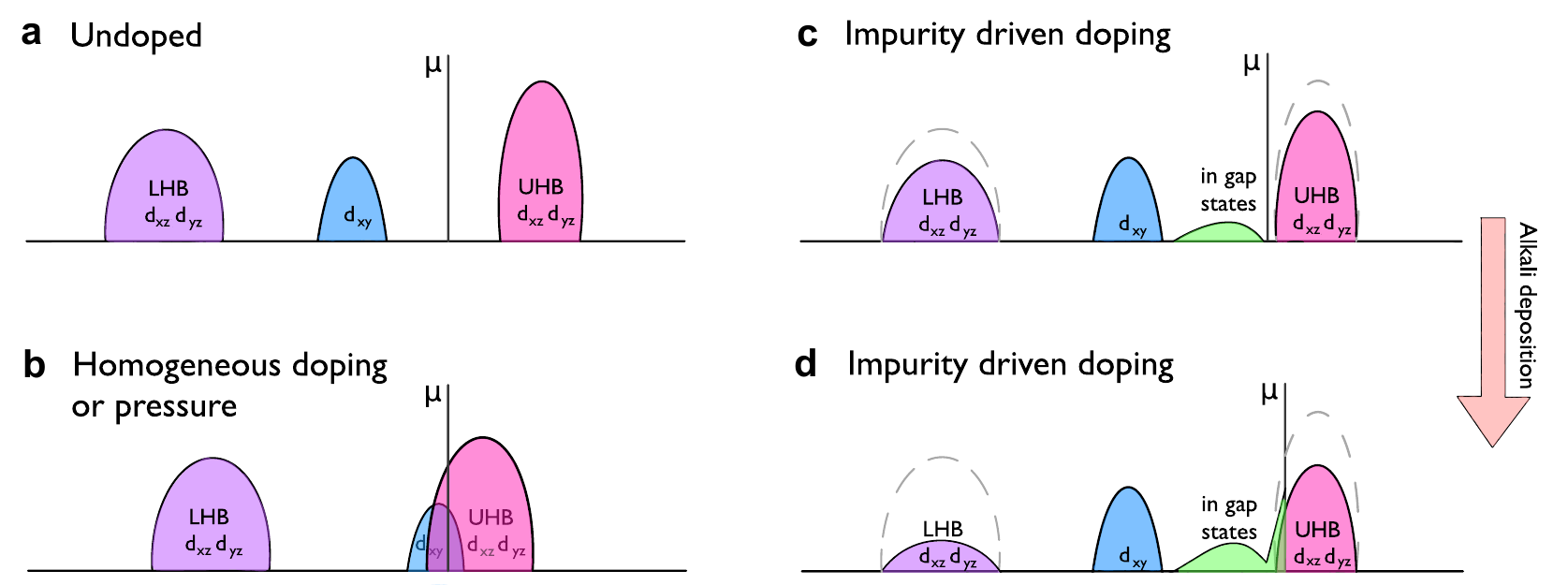}
		\end{center}
		\caption{\textbf{Mott-insulator re-metallization schemes of \CRO.} (a) Schematic representation of the \CRO\ band-Mott insulating structure~\cite{DasPRX2018}. In the short $c$-axis phase, the \dxy\ band is fully occupied whereas the half-filled \dxz,\dyz\ orbitals are Mott insulating forming lower and upper Hubbard bands. (b) Homogeneous bulk doping, and chemical or applied pressure are expected to generate a multi-band metal that involves all the $t_{2g}$ orbitals. (c)  Impurity electron doping, from surface alkali-metal deposition, 
			turns out to generate in-gap states through hybridization with the Mott insulating \dxz,\dyz\ orbitals. The in-gap states 
			emerge in conjunction with partial spectral weight suppression of 
			the Hubbard bands (indicated by dashed grey lines). (d) Re-metallization upon crossing a certain threshold of impurity doping level. Due to the orbital-selective hybridization between surface dopants and the high-energy incoherent excitations, a single band metal is formed. 
		}	
		\label{schematic}
	\end{figure*}

	Here we demonstrate -- using photoemission spectroscopy -- that alkali add-on atoms on Mott insulating Ca$_2$RuO$_4$ generates such an intertwined 
	quantum well state. 
	Independent of the chosen alkali-metal element (K, Rb, Cs), we observe  a depletion of the lower-Hubbard-band spectral weight and an evolution of the Ru 
	core-level states as a function of doping. 
	Eventually, a single-band metal emerges as a result of the interaction between the alkali-metal dopants and the \CRO\ substrate [see Figs.~\ref{schematic}(c) and (d)]. Our work reveals 
	a new type of orbital-selective surface  metal-insulator transition induced through in-gap state formation generated by orbital hybridization between surface dopants and the Mott insulating substrate (schematically illustrated in Fig.~1).  \\[2mm]
	

	\begin{figure*}[ht!]
		\begin{center}
			\includegraphics[width=\textwidth]{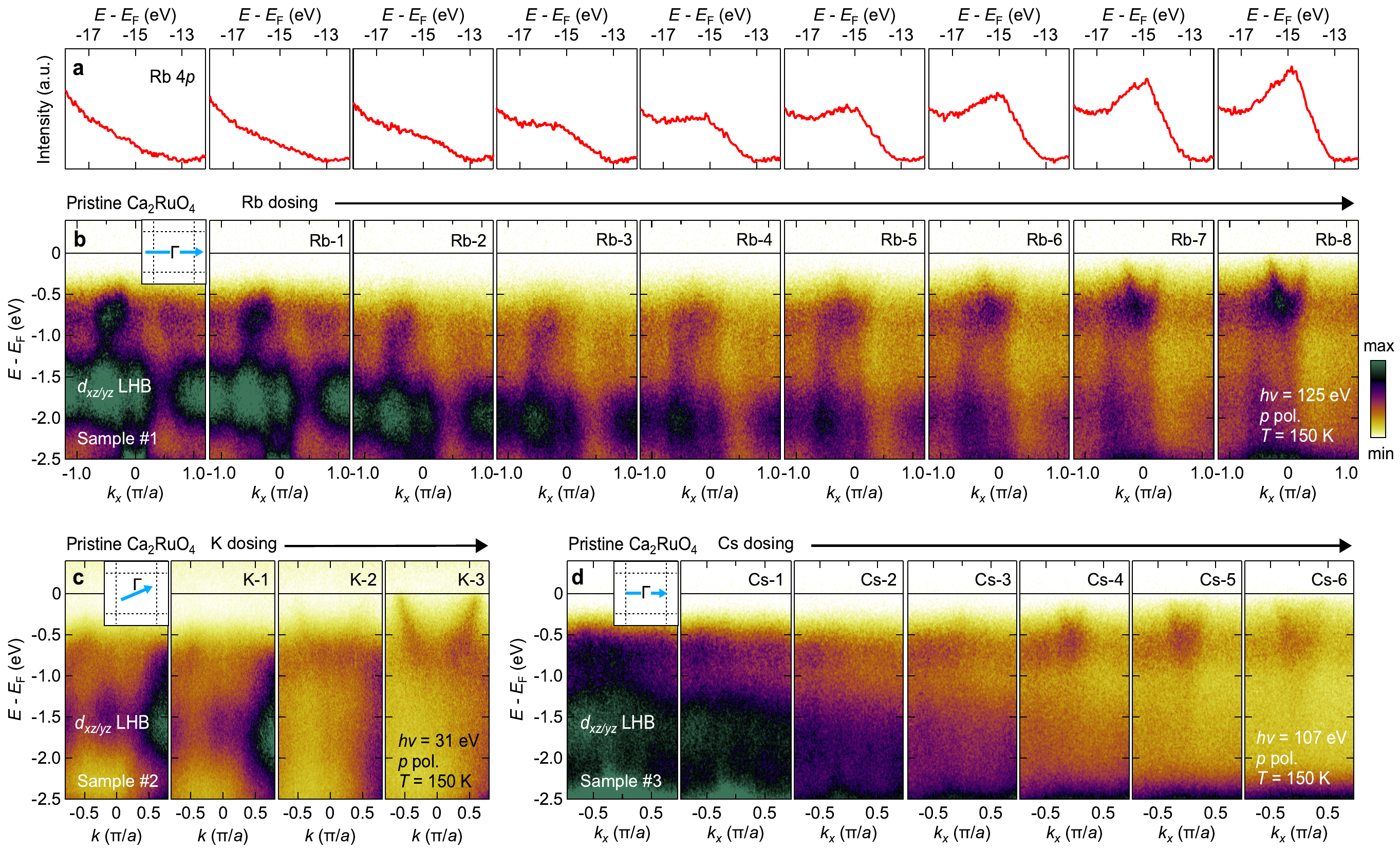}
		\end{center}
		\caption{\textbf{Surface electronic structure evolution of \CRO\ by alkali-metal dosing.} (a),(b) Rb $4p$ energy distribution curves (EDCs) and valence-band energy distribution maps (EDMs), respectively, recorded at $T = 150$~K with dosing Rb in incremental steps. $p$-polarized 125-eV incident light is used. The momentum cut is indicated in the inset of the leftmost panel. (c),(d) EDMs recorded with dosing K and Cs using $p$-polarized light of $h\nu=31$ and 107~eV, respectively. Similar changes as the Rb case are observed.
		}	
		\label{fig:fig1}
	\end{figure*}

	\textbf{Results}\\
	
	\textbf{Electronic-structure evolution by alkali-metal dosing:}
	Rb $4p$ core level spectroscopy and electronic band structure of \CRO\ single crystals as a function of alkali-metal deposition are shown in Figs.~\ref{fig:fig1}(a) and (b), respectively. Angle-resolved photoemission spectroscopy (ARPES) spectra are recorded along the $\Gamma$--M direction. Before alkali-metal deposition,
	the electronic structure -- shown in Fig.~2b, left panel -- is consistent with previous reports~\cite{SutterNatCom2017,RiccoNatCommun18}. 
	Around the Brillouin zone center, dispersive bands (2.5 -- 0.5~eV) are observed and the non-dispersive lower Hubbard band (LHB) is located about $\sim 1.7$~eV below the Fermi level (\EF). Previously, the dispersive bands have been assigned predominately to the \dxy\ orbital and the LHB to the \dxz/\dyz\ orbitals~\cite{SutterNatCom2017,RiccoNatCommun18}. Upon dosing Rb, the whole structure is shifted downwards [Fig.~\ref{fig:fig1}(b)] and a Rb $4p$ core level peak develops [Fig.~\ref{fig:fig1}(a)].
	By further dosing Rb, an in-gap state with an electron-like dispersion evolves from its band bottom. Whereas the low-energy spectral weight is initially negligible, the band extends with the increase of the Rb amount and finally 
	produces finite spectral weight at \EF\ (See Supplementary Fig.~1). These changes occur qualitatively in the same manner irrespective of the choice of alkali-metal dopants -- see Figs.~\ref{fig:fig1}(c) and (d) for the case of K and Cs dosing. 

	\begin{figure*}[ht!]
		\begin{center}
			\includegraphics[width=0.95\textwidth]{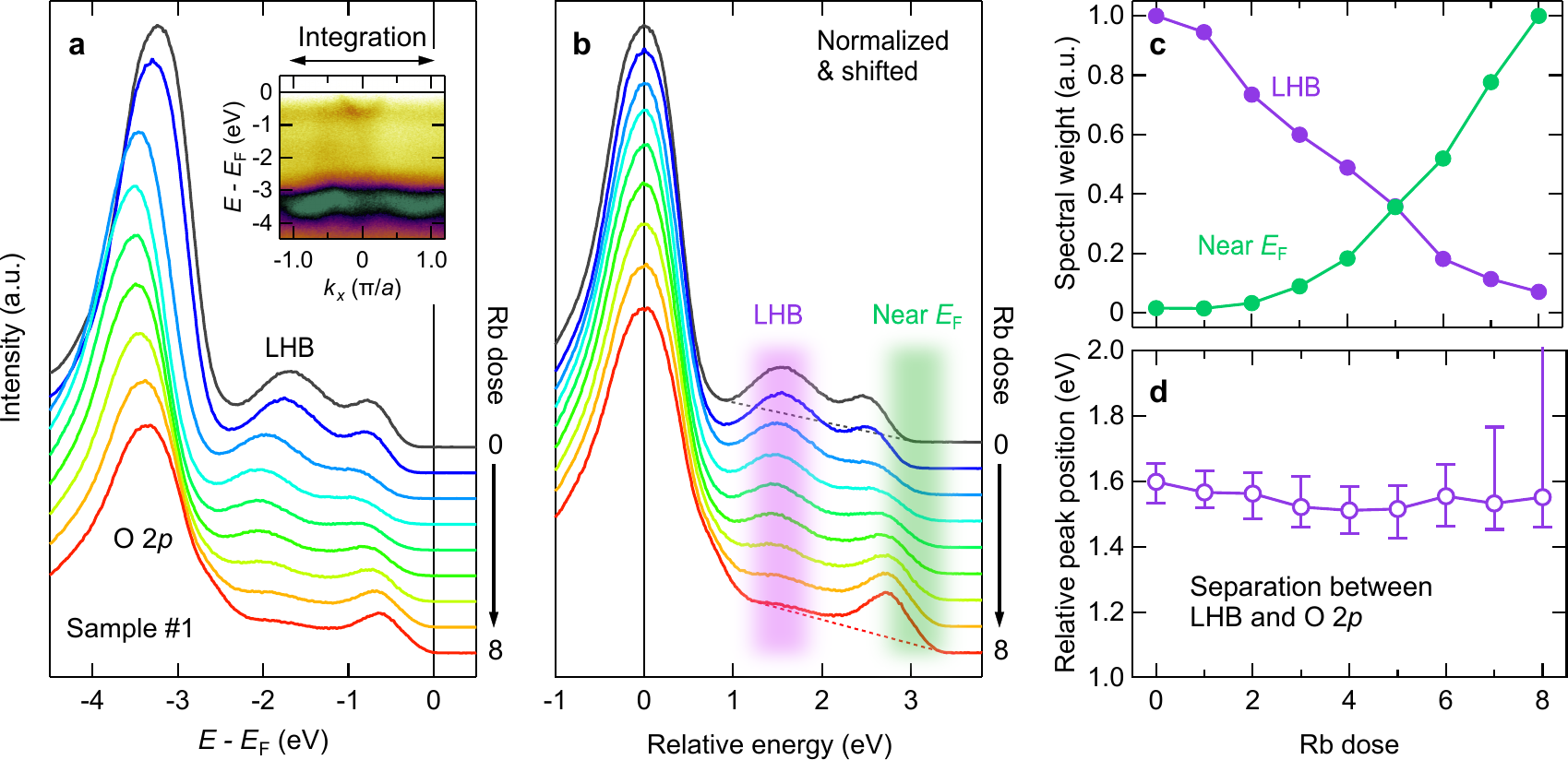}
		\end{center}
		\caption{\textbf{Surface Mott breakdown in \CRO.} (a) Valence-band EDCs versus energy relative to \EF\ plotted in the order of Rb deposition levels. Spectra are given arbitrary vertical offsets for a visibility purpose. The inset indicates the momentum window for integration. (b) Valence band EDCs aligned to the O $2p$ peak position and normalized to the total intensity in the displayed energy region. (c) Spectral weight of LHB and the near-\EF\ part integrated within the magenta and green shaded regions, respectively, in panel (b). The LHB spectral weight is estimated after subtracting a tangential linear background shown by dotted lines in  (b). The weight has been normalized to the maximum values. (d) The position of LHB with respect to the O $2p$ peak position plotted as a function of the Rb deposition sequence. The error bar is determined from intensity variation around the peak surpassing the noise level.
		}	
		\label{fig:fig2}
	\end{figure*}

	\vspace{5mm}
	\textbf{Surface Mott breakdown in \CRO:}
	Momentum-integrated energy distribution curves (EDCs) are plotted in Fig.~\ref{fig:fig2}(a). Besides Ru $4d$-derived states including LHB, O $2p$ states are identified at $\sim 3.2$~eV below \EF~\cite{SutterNatCom2017}. Upon Rb dosing, the O $2p$ peak moves to higher binding energy until the shift saturates at $\sim 0.3$~eV, followed by a slight shift backward. While Ru ions could change their valency when doped with electrons, oxygen ions should remain chemically unperturbed. The O $2p$ peak position thus serves as a measure of the chemical-potential shift as demonstrated by previous studies of oxide materials~\cite{HuNatCommun2021,ShenPRL2004}. 
	In Fig.~\ref{fig:fig2}(b), we align the O $2p$ peak position to compensate for the chemical-potential shift
	and unravel in this fashion the intrinsic lower-energy structures.
	The total spectral intensities, within the displayed energy window, are normalized to eliminate attenuation effects from the Rb overlayer.
	After these data treatments, a prominent decay of LHB intensity (magenta-shaded region) is observed concomitantly with the growth of near-\EF\ spectral weight (green-shaded
	region) -- see Fig.~\ref{fig:fig2}(c). In contrast to the drastic changes in the spectral weight, the position of the LHB 
	is essentially independent of alkali dosing [Fig.~\ref{fig:fig2}(d)]. This suggests that the size of the Mott gap is unaffected by alkali-metal dosing even though the chemical potential moves inside the gap. Instead of the shrinkage of the gap size, the collapse of the Mott gap is driven by the spectral-weight loss in the Hubbard bands. A similar type of the Mott transition with the fixed Mott-gap size has recently been demonstrated for the cuprate Ca$_3$Cu$_2$O$_4$Cl$_2$ 
	upon alkali-metal dosing~\cite{HuNatCommun2021}.

	\begin{figure}[ht!]
		\begin{center}
			\includegraphics[width=0.48\textwidth]{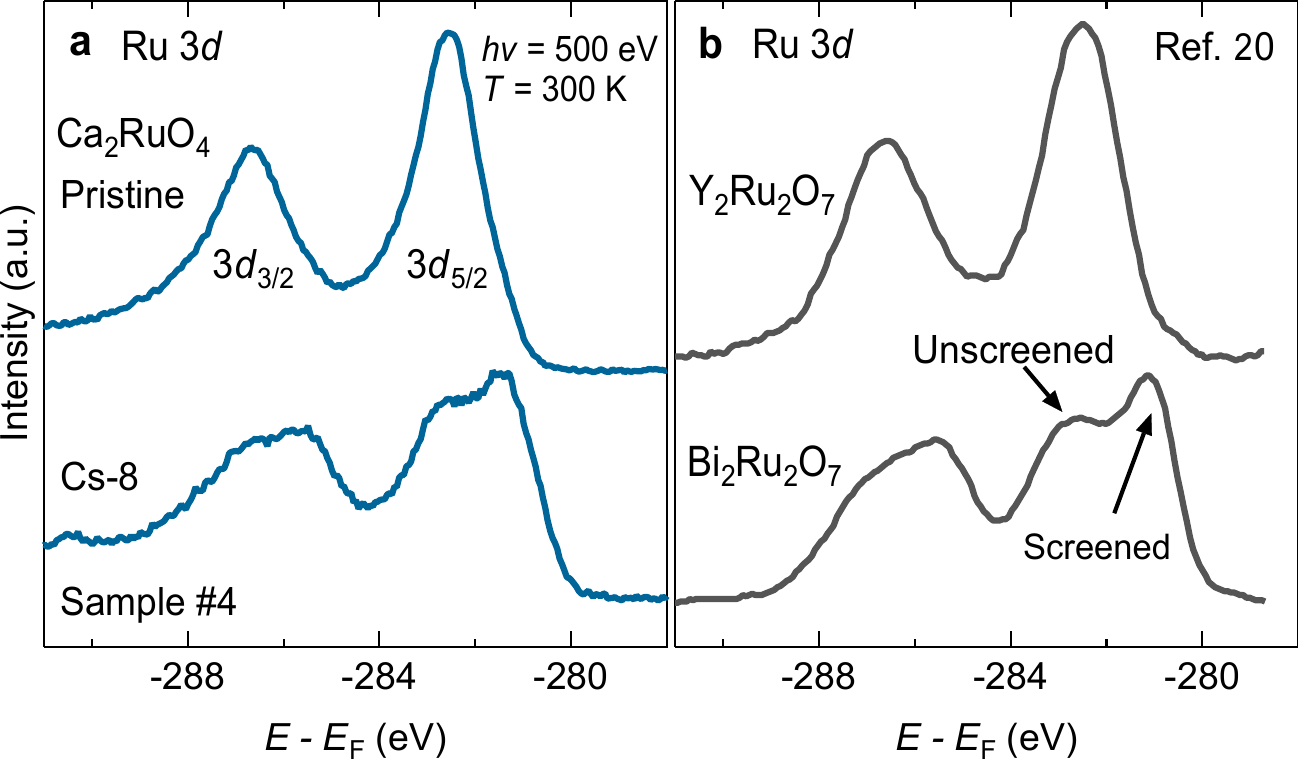}
		\end{center}
		\caption{\textbf{Evolution of core levels.} (a) Ru $3d$ XPS spectra of \CRO\ measured before and after dosing Cs at 300~K. (b) Ru $3d$ XPS spectra of Mott-insulating Y$_2$Ru$_2$O$_7$ and metallic Bi$_2$Ru$_2$O$_7$ from Ref.~\onlinecite{KimPRL2004} with the binding energy aligned to that of \CRO.}
		\label{Ru3d}
	\end{figure}

	\vspace{5mm}
	\textbf{Core-level structure:}
	The Ru $3d$ core levels, probed by x-ray photoemission spectroscopy (XPS), provide insights into the dosing-induced metallic surface state. In Mott insulating \CRO\ and Y$_2$Ru$_2$O$_7$, the Ru $3d$ peak is composed of a single set of spin-orbit-split peaks ($3d_{5/2}$ and $3d_{3/2}$) as shown in Figs.~\ref{Ru3d}(a) and (b). 
	Isovalent Bi substitution for Y~\cite{KimPRL2004}  drives a band-width-controlled Mott transition, and the resulting metallic state of Bi$_2$Ru$_2$O$_7$ yields 
	splittings within the $3d_{5/2}$ and $3d_{3/2}$ levels [Fig.~\ref{Ru3d}(b)]. 
	It has been proposed that the low-energy peak stems from a final state where core holes are screened by conduction electrons~\cite{KimPRL2004}. Upon Cs dosing of \CRO, a very similar splitting of the core levels is observed [Fig.~\ref{Ru3d}(a)] -- suggesting the emergence of conduction electrons with Ru character. 

	\begin{figure*}
		\begin{center}
			\includegraphics[width=0.99\textwidth]{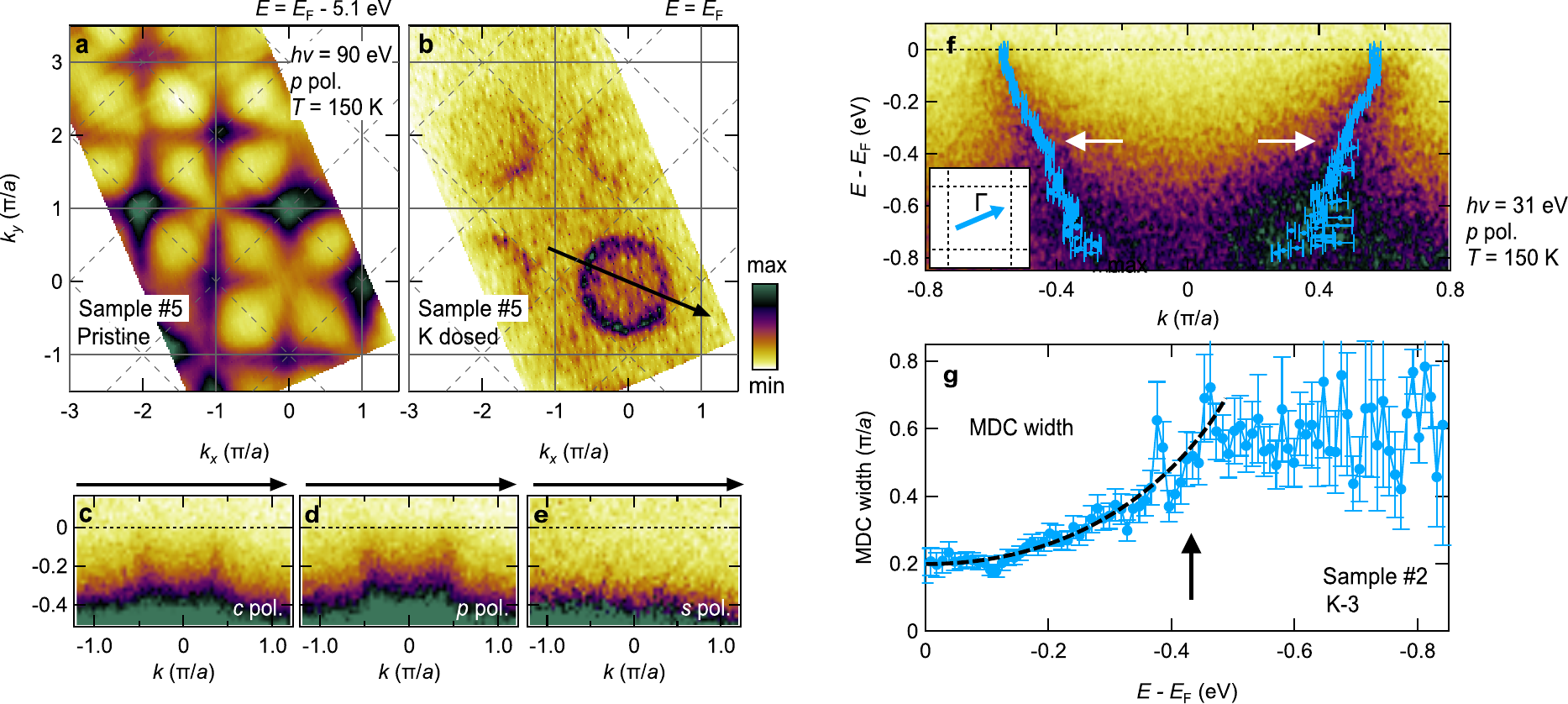}
		\end{center}
		\caption{\textbf{Metallic surface state of alkali-metal dosed \CRO.} (a) Constant-energy map of pristine \CRO\ at $E = E_\mathrm{F} - 5.1$~eV recorded with $h\nu = 90$~eV, $p$-polarized photons at $T = 150$~K. The oxygen bands draw a periodic structure. Overlaid grey solid and dashed squares represent Brillouin zones in the tetragonal and orthorhombic notations, respectively. (b) Fermi surface after K dosing for 10~minutes at 6.5~Ampere). (c)-(e) EDMs recorded with circular, $p$, and $s$ polarization, respectively, along the cut indicated in (b). The band becomes indiscernible when measured with $s$-polarized light. (f) EDM of the K-dosed state [identical to K-3 in Fig.~\ref{fig:fig1}(c)] overlaid with momentum-distribution-curve (MDC) peak positions, recorded with $h\nu = 31$~eV, $p$-polarized photons along the cut shown in the inset. (g) MDC full width at half maximum averaged over the two branches. The dashed curve indicates the low-energy dependence to guide the eye. Kinks in the MDC peak position and width are indicated by arrows. Error bars are based on the standard deviation of the fitting.}
		\label{fig:fig3}
	\end{figure*}

	\vspace{5mm}
	\textbf{Surface metallic state:}
	Having tracked in detail the evolution of the electronic structure by alkali-metal dosing, we now focus on the character of the created metallic state. To embed the momentum-resolved photoemission signal into the Brillouin zone of \CRO, we use the constant-energy map intersecting oxygen bands~\cite{SutterNatCom2017} -- see Fig.~\ref{fig:fig3}(a). Then, Fermi surface after metallization is mapped out in the same momentum regions. The resulting map [Fig.~\ref{fig:fig3}(b)] reveals circular Fermi surfaces in accordance with the tetragonal Brillouin zone. 
	The bulk crystal structure of \CRO\ is orthorhombic.
	Numerous ARPES studies have accumulated evidence that the potential of orthorhombic distortion is strong enough to cause band folding~\cite{DamascelliPRL00,TamaiPRL2008,LiuPRB2018,SutterPRB19,HorionpjQM2021}. 
	The absence of band folding
	suggests that the metallic surface state is not strictly confined in the Ca$_2$RuO$_4$ crystal. 
	The spectral intensity of the electron-like band composing the Fermi surface is strongly suppressed with $s$-polarized light. In contrast, the band is visible with $p$-polarized light irrespective of the azimuthal angle of the mirror plane [see also Figs.~\ref{fig:fig1}(b) and (c)]. The band therefore possesses in-plane even character. As shown in Fig.~\ref{fig:fig3}(f), the metallic band exhibits a kink (sudden change of band velocity) at $\sim 0.4$~eV below \EF. This is also evidenced by a saturation of the   momentum distribution curve (MDC) linewidth [Fig.~\ref{fig:fig3}(g)].

	\vspace{5mm}
	\textbf{Discussion}\\
	Our main observation is a metallization of \CRO\ upon application of alkali-metal atoms. 
	A central question is whether this quantum well state is a hybrid state between alkali metals and \CRO. 
	We address this question by inspecting: (i) the spectral weight of the LHB, (ii) the Ru-core levels, and (iii) the self-energy effects of the induced quantum well state.

(i) Mono-, bi-, and tri-layer alkali-metal deposition have been reported~\cite{KimScience2014,KyungnpjQM2021} on \SIO\ and \SRO. Due to the large inelastic mean free path of alkali metals~\cite{SmithSS1993}, bulk bands are observed through the alkali-metal layers in both cases.
The observation of drastic LHB suppression (Fig.~\ref{fig:fig2}) is thus intrinsic and not an artifact of an alkali-metal overlayer.
With few exceptions~\cite{ZhouPRB2016}, spectral weight suppression of the Hubbard bands is  associated with quasi-particle formation near the chemical potential~\cite{EskesPRL1991,HuNatCommun2021}. This is consistent with our observation of a fading Hubbard band being replaced by a valence band as a function of alkali-metal dosing.

(ii) Also the Ru core level [Fig.~\ref{Ru3d}(a)] is modified by alkali-metal deposition. The simplest possibility of lower-energy satellite in the Ru $3d$ peak is a Ru$^{3+}$ component created by doping Ru$^{4+}$ with an electron. However, the intense low-energy peak translates into more than 0.6 electrons doped per Ru atom. This value is unrealistic as previous alkali-metal adsorption studies on oxides found at most $\sim 0.15$ electrons doped per atom~\cite{KimScience2014,YukawaPRB2018,KyungnpjQM2021,HuNatCommun2021}. Instead, such intense low-energy satellite peak, which appears upon metallization of Mott insulators like cuprates~\cite{TaguchiPRL2005,HorioPRL2018_2} and ruthenates~\cite{Cox1983,KimPRL2004}, has been associated with a final state where core holes are efficiently screened by conduction electrons. The 
change induced by alkali-metal dosing [Fig.~\ref{Ru3d}(a)] thus suggests the emergence of itinerant electrons with, at least partially, the Ru $3d$ character. We note that
the linewidth of other core levels such as Ca $3p$  is essentially insensitive to dosing of alkali atoms [see Supplementary Fig.~2] 
suggesting that the Ru $3d$ transformation is intrinsic.

\begin{figure*}[ht!]
	\begin{center}
		\includegraphics[width=0.95\textwidth]{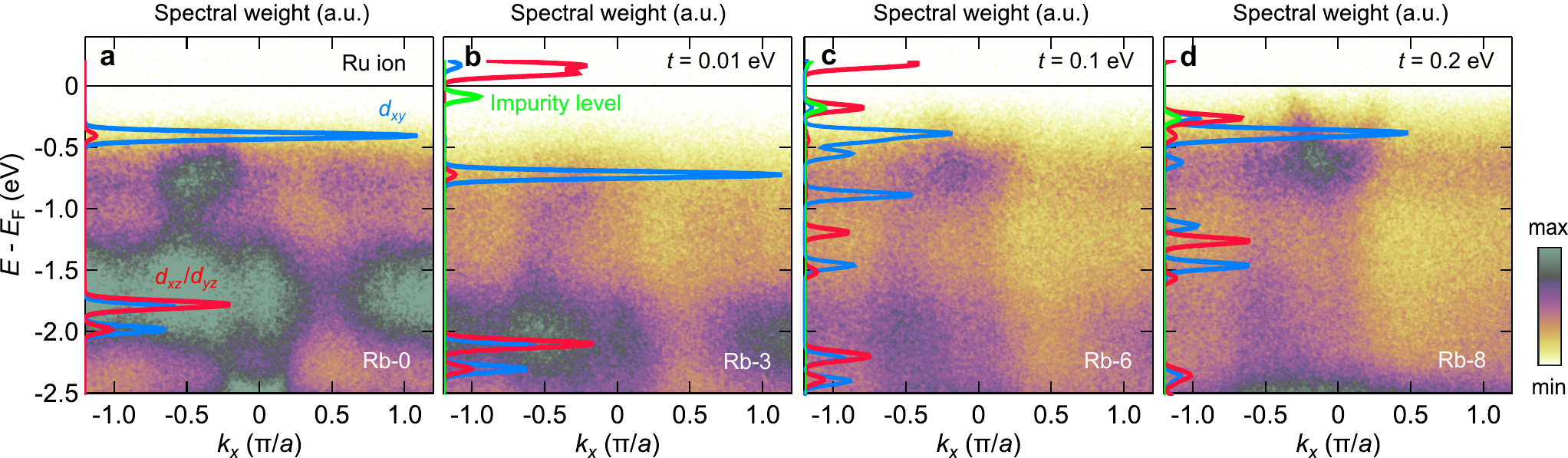}
	\end{center}
	\caption{\textbf{Cluster analysis of \CRO\ interacting with an alkali-metal impurity level.} (a)-(d) Spectral function calculated for Ru ion, $t=0.01$~eV, $t=0.1$~eV, and $t=0.2$~eV, respectively. Superimposed are energy distribution maps from Fig.~\ref{fig:fig1}(b) with Rb deposition sequences as indicated.
	}	
	\label{fig:fig5}
\end{figure*}

(iii)
Finally, the kink witnessed both in the band dispersion [Fig.~\ref{fig:fig3}(f)] and MDC linewidth [Fig.~\ref{fig:fig3}(g)] of the metallic band suggests strong self-energy effects. The energy scale of ~0.4~eV is incompatible with electron-phonon interactions and rather points to electron-electron interactions. In fact, such high-energy kinks have been widely observed in strongly correlated systems like cuprates~\cite{XiePRL2007,ChangPRB2007} and ruthenates~\cite{IwasawaPRL2012}, and interpreted as a manifestation of many-body self-energy effects. The kink, therefore, suggests substantial electron correlations induced by substrate Ru orbitals. Such correlation effects are not expected (or reported) for a purely alkali-metal quantum-well states~\cite{EknapakulPRB2016}.

Based on observations (i)-(iii), we conclude that the 
induced metallic state involves an interaction between alkali-metal atoms and \CRO.

So far, bulk electron doping of \CRO\ has been achieved only 
by substituting La or Pr for Ca~\cite{FukazawaJPSJ2000,CaoPRB2000,Pincini2019}. This substitution inevitably involves chemical pressure. 
As 
a similar metal-insulator transition is observed with isovalent Sr substitutions~\cite{NakatsujiPRL00}, chemical pressure rather than doping is most likely the metallization mechanism of La/Pr doping. In all cases, the resulting metallic state comprises  multiple Fermi surfaces -- composed of almost evenly filled \dxz, \dyz, and \dxy\ orbitals~\cite{RiccoNatCommun18,SutterPRB19}. 
These 
electronic structures are 
realistically captured  
by dynamical mean-fied theory (DMFT) calculations, which take into account both electronic structure and strong correlations (self-energy) effects~\cite{RiccoNatCommun18,SutterPRB19}. In the present surface-dosed case, it is unclear whether the transition involves $c$-axis changes. We therefore performed DMFT calculations for the short $c$-axis phase to evaluate genuine electron-doping effects. As shown in Supplementary Fig.~3, electron doping again leads to the formation of multiple Fermi surface sheets (see also Supplementary Note~1). 
This demonstrates that alkali metal dosing does not correspond to the standard theoretical description of homogeneous electron doping. 
Surface re-metallization due to 
Coulomb screening 
is also expected to yield a multiband metal.

Inspection of the initial stage of the band-structure evolution by alkali-metal dosing  [Figs.~\ref{fig:fig1}(b)-(d)] indicates no detectable spectral weight accumulation at \EF. This is in direct contrast to electron doping of Mott insulating cuprates~\cite{ArmitagePRL2002,HuNatCommun2021}, and is an indication that the charge transfer of the alkali valence electrons to the UHB band is not complete. This result 
suggests a covalent nature of bonding 
with partial charge transfer to the Ru bands. 
This indicates that  alkali-metal $s$ electrons 
are 
not directly injected into the UHB. Instead, they reside on localized impurity states with pinned chemical potential within the gapped region. Such impurity states can hybridize with the Ru $d$ orbitals provided sufficient spacial overlap and symmetry compatibility. Linear combination of  \dxz\ and \dyz\ orbitals (with even parity) is the most obvious candidate for such hybridization. The formation of interfacial bonding states between surface-dosed alkali-metal $s$ and transition-metal $d$ orbitals has been frequent observation in previous photoemission studies~\cite{SoukiassianPRB1985,AkikuboTSF2016,OzawaSS1997,OzawaSS1999,ZhangPRB2016}.

We  therefore evaluated the orbital-dependent spectral function via exact diagonalization of a cluster made of one Ru site and one impurity level. The model consists of the local interaction terms at the Ru and the impurity sites, and the kinetic term among the \dxz/\dyz\ and the impurity orbitals $t$. The following choices are made for the parameters: $U=2$~eV, $J=0.5$~eV, $\Delta_\mathrm{CF}=0.3$~eV, $\lambda=0.075$~eV, $U_\mathrm{I}=1$~eV~\cite{DasPRX2018,SutterNatCom2017}, where $U$ is the Coulomb interaction between Ru 4$d$ electrons, $J$ is the Hund's coupling, $\Delta_\mathrm{CF}$ is crystal-field splitting within the $t_{2g}$ sector, $\lambda$ is spin-orbit coupling, and $U_\mathrm{I}$ is the Coulomb interaction within the impurity level. We assume that the impurity level lies close to the bottom of the UHB and we mimic the increase of the doping concentration by increasing the parameter $t$, allowing more hybridization between the Ru \dxz/\dyz\ and impurity level. As shown in Figs.~\ref{fig:fig5}(a) and (b), the initial change with small $t$ is the shift of chemical potential to accommodate the impurity level. Upon increasing hybridization, the spectral weight of the \dxz\,\dyz\  located at $E \sim E_\mathrm{F}-2$~eV is significantly suppressed and instead a new state with a mixed character of Ru \dxz/\dyz\ and impurity $s$ appears near \EF\ [Fig.~\ref{fig:fig5}(c)]. Further increase of $t$ results in accumulating the spectral weight of this bonding state [Fig.~\ref{fig:fig5}(d)]. The overall changes are in good agreement with the experiment. 
We can provide a qualitative interpretation of the cluster calculations by analyzing the multiplet eigenstates and electronic transitions for the Ru and the alkali impurity state at the ionic level. A detailed analysis is reported in Supplementary Figs.~4, 5, Supplementary Notes~2 and 3. The ground state is built up as a quantum superposition of local $|d,s\rangle$  configurations with a fully occupied \dxy\ orbital, due to the extreme flattening of the Ru-O octahedra. Assuming a selective hybridization among the $s$ level and \dxz,\dyz\ doublet, this superposition involves $|d^4, s^1\rangle$ and  $|d^5,\underline{s} \rangle$ states. Thus, two spectral features can be obtained via the removal of one electron from the \dxz,\dyz\ doublet, which arise from an electronic transition to the $ |d^3\ s^1\rangle$  and  $|d^4,\underline{s} \rangle$ state, respectively.  While the former contributes to the weight of the LHB,  the latter corresponds to the in-gap states and is expected to increase its spectral weight as the deposition sequence goes on.

The effects of such orbital-selective hybridization mechanism on the extended system are schematized in Fig.~\ref{schematic}(d). Here we show that the Mott breakdown takes place due to the progressive depletion of the  \dxz,\dyz\ LHB
and consequent filling of the in gap impurity driven states. It is worthwhile to notice that such mechanism has side effects also on the \dxy\ band. Due to the covalent nature of the ground state, the spectral weight associated to the \dxy\ removal states spreads out between several allowed electronic transitions at slightly higher binding energies (See Supplementary Figs.~4, 5, and Supplementary Note~3 for details).

Finally, we notice that the orbital-selective hybridization can naturally lead to the formation of a single-sheet Fermi surface that is not reproduced by the homogeneous doping picture. The symmetry analysis of the metallic state (Fig.~\ref{fig:fig3}) is also compatible with the in-plane even character of the bonding state. 

\begin{figure}[ht!]
\begin{center}
	\includegraphics[width=0.47\textwidth]{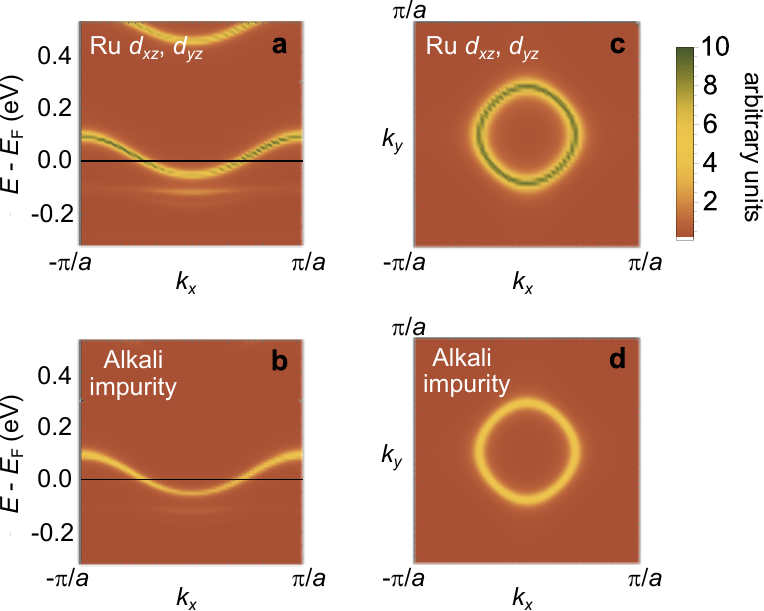}
\end{center}
\caption{Electronic dispersion at $k_y$=0 by varying $k_x$ in the range [-$\pi$,$\pi$] for the bands (a) arising from the Ru $d_{xz}$,$d_{yz}$ orbitals and (b) from the impurity $s$-state at the alkali site. Black lines identify the fermi level. Fermi pocket of the (c) Ru $d_{xz}$,$d_{yz}$  bands and (d) impurity band. The corresponding Fermi lines indicate the occurrence of an electron pocket centered at $\Gamma$. The spectral weight of the pocket is made of hybridized $d_{xz}$,$d_{yz}$  and s bands. The computation refers to an electronic configuration with vanishing direct hybridization between the impurity states on nearest neighbors. The dispersion of the impurity $s$-state is yielded through the local hybridization with the Ru $d$-orbitals and the Ru-Ru effective charge transfer.}
\label{fig:figCPT}
\end{figure}

To explore whether our orbital-selective hybridization theoretical model is compatible with the observation of a single Fermi surface sheet with free-electron like dispersion, we combined the cluster calculation with cluster perturbation theory (CPT). The latter uses exact diagonalization of small clusters to construct a strong-coupling perturbation theory for the lattice problem \cite{Senechal2002}  (see Supplementary Note 4).  Even though this approach represents a simplification of the complex physics characterizing the large dopant regime, it allows to determine the Fermi surfaces to be compared to our experimental results. In particular, we are interested in the evolution with alkali content of the low-energy features corresponding to the in-gap states. In the atomic regime, as demonstrated in the cluster calculation, the in-gap localized states are mainly made of $s$- and $d_{xz}/d_{yz}$ states with a relative charge distribution dictated by the hybridization and the multiplet configurations at the Ru sites. These localized states can overlap along the Ru-O-Ru bond directions. In order to understand the formation of the Fermi pocket as due to the hybridization of the $s$-states with the Ru $d$ bands, it is particularly instructive to consider the limit with no direct overlap between the $s$- states. In this case, the effective mass of the $s$-state is due to the hybridization through the mixing with the Ru $d$-states across the Mott gap. As shown in Fig. \ref{fig:figCPT}, we observe that the impurity level acquires an effective mass with a dispersion that substantially follows that one of the  $d_{xz,dyz}$  bands, due to the local hybridization. 
This result is confirming that our orbital selective model supports the formation of a  Fermi pocket, since the impurity states can get indirectly connected through the $d$-bands in the insulating Mott phase. The analysis has been performed for a representative case for the set of parameters used for the single site cluster calculation. Small variations do not affect the qualitative character of the Fermi pocket. The overall outcome is compatible with our experimental observation. We conclude that this scenario provides a novel type of surface Mott-insulator to metal transition realized through chemical doping.

Due to the reduced coordination at surfaces, correlated systems have the opposite tendency, namely that of spontaneously forming a skin with suppressed conductivity. This has been observed and discussed in various oxides such as vanadates~\cite{SekiyamaPRL2004,RodolakisPRL2009,YoshimatsuPRL2010}, cuprates~\cite{TaguchiPRL2005}, and ruthenates~\cite{TakizawaPRB2005,PanaccioneNJP2011}. The present work demonstrates the opposite case where dosed alkali metals increase surface hopping channels and produces a metallic skin 
on the Ca$_2$RuO$_4$ Mott insulating state. Although the multiband Mott insulating state of Ca$_2$RuO$_4$ is rather unique in nature, similar physics could be realized in other systems. Typical ingredients would be quasi-two dimensional materials with Mott bands composed of inter-layer directed orbitals ($p_z$, $d_z$ or $d_{xz}$) and moderate electron correlation. These criteria are uniquely satisfied in Ca$_2$RuO$_4$. The Mott bands have predominately 4$d_{xz/yz}$ character and the moderate correlation strength makes  hybridization with deposited alkali-metal electrons possible.   

\vspace{5mm}
\textbf{Methods}\\
\textbf{Sample preparation and photoemission experiments:} High-quality \CRO\ single crystals were grown by the flux-feeding floating-zone method~\cite{FukazawaPhysB00,snakatsujiJSSCHEM2001}. ARPES measurements were carried out at the MAESTRO and I05 beamlines at Advanced Light Source and Diamond Light Source, respectively, at 150~K in the insulating short $c$-axis phase. Photon energy was varied in the range of 30--130~eV with energy resolution better than 50~meV. Presented ARPES data are recorded either using $s$-, $p$-, or circularly polarized light, with $s$ and $p$ denoting odd and even parity with respect to the photoemission mirror plane, respectively. XPS measurements were conducted at BL07LSU of SPring-8 at 300~K and 500~eV of incident photon energy. Samples \#1--\#3 and \#5 were measured with ARPES and sample \#4 with XPS. For all the photoemission measurements, samples were cleaved \textit{in-situ} using the top-post method. SAES Getter dispensers were used to evaporate K, Rb, or Cs onto the \CRO\ surface in incremental steps. Unless otherwise stated, one dose corresponds to evaporation of K, Rb, and Cs for respectively 40, 30, and 30 seconds with a filament current of 6.6, 6.4, and 8.3~Ampere.
Note that Cs dosing for the ARPES and XPS measurements were done at different instruments and hence the dose unit is not exactly equivalent.
No detectable charging was observed when varying the photon flux as long as the temperature was kept above 150~K. The tetragonal notation with $a=3.80$~\AA\ is used to display ARPES data.\\

\textbf{Data availability} \\
The data that support the findings of this study are available from the corresponding author upon reasonable request. \\

\textbf{Acknowledgements}\\
Fruitful discussion with M.~Grioni is greatfully acknowledged. M.~Horio, D.S., C.G.F., C.E.M., and J.C. acknowledge support by the Swiss National Science Foundation. M.~Horio and T.W. are supported by Grants-in-aid from the Japan Society of the Promotion of Science (JSPS) (grant No.~21K13872). M.K. was supported by KIAS Individual Grants (CG083501). S.M. acknowledges support by the Swiss National Science Foundation under grant no.~P300P2-171221. Y.S. is funded by the Swedish Research Council (VR) with a Starting Grant (Dnr. 2017-05078). 
G.S. acknowledges the hospitality of the Center for Computational Quantum Physics at the Flatiron Institute, a division of the Simons Foundation, as well as funding support from the Deutsche Forschungsgemeinschaft (DFG, German Research Foundation) under Germany's Excellence Strategy through the W\"urzburg-Dresden Cluster of Excellence on Complexity and Topology in Quantum Matter ct.qmat (EXC 2147, Project ID 390858490) as well as through the Collaborative Research Center SFB 1170 ToCoTronics (Project ID 258499086).
This research used resources of the Advanced Light Source, which is a DOE Office of Science User Facility under contract no. DE-AC02-05CH11231. We acknowledge Diamond Light Source for time at beamline I05 under proposal SI10550. We are grateful to the CPHT computer support team for the DFT+DMFT computation.\\

\textbf{Author contributions}\\
V.G., R.F., and A.V. grew the \CRO\ single crystals. M. Horio, D.S, C.G.F., C.E.M., S.M., Y.S., G.G., and J.C. carried out the ARPES experiments. The ARPES data were analyzed by M.Horio and D.S. M.Horio and T.W. performed the XPS experiments and analyzed the data. Photoemission beamlines were developed and maintained by M. Hoesch, T.K.K., S.M., C.J., A.B., E.R., and I.M. DMFT calculations were conducted by M.K., A.G., and G.S. Cluster-diagonalization calculations were carried out by F.F. and M.C. M.H., F.F., M.C., and J.C. wrote the manuscript with inputs from other authors. \\

\textbf{Competing interests} \\
The authors declare no competing interests.\\

\textbf{Additional information}\\
Correspondence to: M.~Horio (mhorio@issp.u-tokyo.ac.jp) and F.~Forte (filomena.forte@spin.cnr.it).


\begin{thebibliography}{52}%
\makeatletter
\providecommand \@ifxundefined [1]{%
\@ifx{#1\undefined}
}%
\providecommand \@ifnum [1]{%
\ifnum #1\expandafter \@firstoftwo
\else \expandafter \@secondoftwo
\fi
}%
\providecommand \@ifx [1]{%
\ifx #1\expandafter \@firstoftwo
\else \expandafter \@secondoftwo
\fi
}%
\providecommand \natexlab [1]{#1}%
\providecommand \enquote  [1]{``#1''}%
\providecommand \bibnamefont  [1]{#1}%
\providecommand \bibfnamefont [1]{#1}%
\providecommand \citenamefont [1]{#1}%
\providecommand \href@noop [0]{\@secondoftwo}%
\providecommand \href [0]{\begingroup \@sanitize@url \@href}%
\providecommand \@href[1]{\@@startlink{#1}\@@href}%
\providecommand \@@href[1]{\endgroup#1\@@endlink}%
\providecommand \@sanitize@url [0]{\catcode `\\12\catcode `\$12\catcode
`\&12\catcode `\#12\catcode `\^12\catcode `\_12\catcode `\%12\relax}%
\providecommand \@@startlink[1]{}%
\providecommand \@@endlink[0]{}%
\providecommand \url  [0]{\begingroup\@sanitize@url \@url }%
\providecommand \@url [1]{\endgroup\@href {#1}{\urlprefix }}%
\providecommand \urlprefix  [0]{URL }%
\providecommand \Eprint [0]{\href }%
\providecommand \doibase [0]{http://dx.doi.org/}%
\providecommand \selectlanguage [0]{\@gobble}%
\providecommand \bibinfo  [0]{\@secondoftwo}%
\providecommand \bibfield  [0]{\@secondoftwo}%
\providecommand \translation [1]{[#1]}%
\providecommand \BibitemOpen [0]{}%
\providecommand \bibitemStop [0]{}%
\providecommand \bibitemNoStop [0]{.\EOS\space}%
\providecommand \EOS [0]{\spacefactor3000\relax}%
\providecommand \BibitemShut  [1]{\csname bibitem#1\endcsname}%
\let\auto@bib@innerbib\@empty
\bibitem [{\citenamefont {Caviglia}\ \emph {et~al.}(2008)\citenamefont
{Caviglia}, \citenamefont {Gariglio}, \citenamefont {Reyren}, \citenamefont
{Jaccard}, \citenamefont {Schneider}, \citenamefont {Gabay}, \citenamefont
{Thiel}, \citenamefont {Hammerl}, \citenamefont {Mannhart},\ and\
\citenamefont {Triscone}}]{CavigliaNat2008}%
\BibitemOpen
\bibfield  {author} {\bibinfo {author} {\bibfnamefont {A.~D.}\ \bibnamefont
	{Caviglia}}, \bibinfo {author} {\bibfnamefont {S.}~\bibnamefont {Gariglio}},
\bibinfo {author} {\bibfnamefont {N.}~\bibnamefont {Reyren}}, \bibinfo
{author} {\bibfnamefont {D.}~\bibnamefont {Jaccard}}, \bibinfo {author}
{\bibfnamefont {T.}~\bibnamefont {Schneider}}, \bibinfo {author}
{\bibfnamefont {M.}~\bibnamefont {Gabay}}, \bibinfo {author} {\bibfnamefont
	{S.}~\bibnamefont {Thiel}}, \bibinfo {author} {\bibfnamefont
	{G.}~\bibnamefont {Hammerl}}, \bibinfo {author} {\bibfnamefont
	{J.}~\bibnamefont {Mannhart}}, \ and\ \bibinfo {author} {\bibfnamefont
	{J.-M.}\ \bibnamefont {Triscone}},\ }\bibfield  {title} {\enquote {\bibinfo
	{title} {Electric field control of the {LaAlO}3/{SrTiO}3 interface ground
		state},}\ }\href {\doibase 10.1038/nature07576} {\bibfield  {journal}
{\bibinfo  {journal} {Nature}\ }\textbf {\bibinfo {volume} {456}},\ \bibinfo
{pages} {624--627} (\bibinfo {year} {2008})}\BibitemShut {NoStop}%
\bibitem [{\citenamefont {Kim}\ \emph {et~al.}(2016)\citenamefont {Kim},
\citenamefont {Sung}, \citenamefont {Denlinger},\ and\ \citenamefont
{Kim}}]{KimNatPhys2016}%
\BibitemOpen
\bibfield  {author} {\bibinfo {author} {\bibfnamefont {Y.~K.}\ \bibnamefont
	{Kim}}, \bibinfo {author} {\bibfnamefont {N.~H.}\ \bibnamefont {Sung}},
\bibinfo {author} {\bibfnamefont {J.~D.}\ \bibnamefont {Denlinger}}, \ and\
\bibinfo {author} {\bibfnamefont {B.~J.}\ \bibnamefont {Kim}},\ }\bibfield
{title} {\enquote {\bibinfo {title} {Observation of a d-wave gap in
		electron-doped {Sr$_2$IrO$_4$}},}\ }\href {\doibase 10.1038/nphys3503}
		{\bibfield  {journal} {\bibinfo  {journal} {Nat. Phys.}\ }\textbf {\bibinfo
	{volume} {12}},\ \bibinfo {pages} {37--41} (\bibinfo {year}
{2016})}\BibitemShut {NoStop}%
\bibitem [{\citenamefont {Liu}\ \emph {et~al.}(2021)\citenamefont {Liu},
\citenamefont {Yan}, \citenamefont {Jin}, \citenamefont {Ma}, \citenamefont
{Hsiao}, \citenamefont {Lin}, \citenamefont {Bretz-Sullivan}, \citenamefont
{Zhou}, \citenamefont {Pearson}, \citenamefont {Fisher}, \citenamefont
{Jiang}, \citenamefont {Han}, \citenamefont {Zuo}, \citenamefont {Wen},
\citenamefont {Fong}, \citenamefont {Sun}, \citenamefont {Zhou},\ and\
\citenamefont {Bhattacharya}}]{LiuScience2021}%
\BibitemOpen
\bibfield  {author} {\bibinfo {author} {\bibfnamefont {Changjiang}\
	\bibnamefont {Liu}}, \bibinfo {author} {\bibfnamefont {Xi}~\bibnamefont
	{Yan}}, \bibinfo {author} {\bibfnamefont {Dafei}\ \bibnamefont {Jin}},
\bibinfo {author} {\bibfnamefont {Yang}\ \bibnamefont {Ma}}, \bibinfo
{author} {\bibfnamefont {Haw-Wen}\ \bibnamefont {Hsiao}}, \bibinfo {author}
{\bibfnamefont {Yulin}\ \bibnamefont {Lin}}, \bibinfo {author} {\bibfnamefont
	{Terence~M.}\ \bibnamefont {Bretz-Sullivan}}, \bibinfo {author}
{\bibfnamefont {Xianjing}\ \bibnamefont {Zhou}}, \bibinfo {author}
{\bibfnamefont {John}\ \bibnamefont {Pearson}}, \bibinfo {author}
{\bibfnamefont {Brandon}\ \bibnamefont {Fisher}}, \bibinfo {author}
{\bibfnamefont {J.~Samuel}\ \bibnamefont {Jiang}}, \bibinfo {author}
{\bibfnamefont {Wei}\ \bibnamefont {Han}}, \bibinfo {author} {\bibfnamefont
	{Jian-Min}\ \bibnamefont {Zuo}}, \bibinfo {author} {\bibfnamefont {Jianguo}\
	\bibnamefont {Wen}}, \bibinfo {author} {\bibfnamefont {Dillon~D.}\
	\bibnamefont {Fong}}, \bibinfo {author} {\bibfnamefont {Jirong}\ \bibnamefont
	{Sun}}, \bibinfo {author} {\bibfnamefont {Hua}\ \bibnamefont {Zhou}}, \ and\
\bibinfo {author} {\bibfnamefont {Anand}\ \bibnamefont {Bhattacharya}},\
}\bibfield  {title} {\enquote {\bibinfo {title} {Two-dimensional
		superconductivity and anisotropic transport at {KTaO$_3$} (111) interfaces},}\ }\href
		{\doibase 10.1126/science.aba5511} {\bibfield  {journal} {\bibinfo  {journal}
	{Science}\ }\textbf {\bibinfo {volume} {371}},\ \bibinfo {pages} {716--721}
(\bibinfo {year} {2021})}\BibitemShut {NoStop}%
\bibitem [{\citenamefont {Liu}\ \emph {et~al.}(2023)\citenamefont {Liu},
\citenamefont {Zhou}, \citenamefont {Hong}, \citenamefont {Fisher},
\citenamefont {Zheng}, \citenamefont {Pearson}, \citenamefont {Jiang},
\citenamefont {Jin}, \citenamefont {Norman},\ and\ \citenamefont
{Bhattacharya}}]{LiuNatComm2023}%
\BibitemOpen
\bibfield  {author} {\bibinfo {author} {\bibfnamefont {Changjiang}\
	\bibnamefont {Liu}}, \bibinfo {author} {\bibfnamefont {Xianjing}\
	\bibnamefont {Zhou}}, \bibinfo {author} {\bibfnamefont {Deshun}\ \bibnamefont
	{Hong}}, \bibinfo {author} {\bibfnamefont {Brandon}\ \bibnamefont {Fisher}},
\bibinfo {author} {\bibfnamefont {Hong}\ \bibnamefont {Zheng}}, \bibinfo
{author} {\bibfnamefont {John}\ \bibnamefont {Pearson}}, \bibinfo {author}
{\bibfnamefont {Jidong~Samuel}\ \bibnamefont {Jiang}}, \bibinfo {author}
{\bibfnamefont {Dafei}\ \bibnamefont {Jin}}, \bibinfo {author} {\bibfnamefont
	{Michael~R.}\ \bibnamefont {Norman}}, \ and\ \bibinfo {author} {\bibfnamefont
	{Anand}\ \bibnamefont {Bhattacharya}},\ }\bibfield  {title} {\enquote
{\bibinfo {title} {Tunable superconductivity and its origin at {KTaO}$_3$
		interfaces},}\ }\href {\doibase 10.1038/s41467-023-36309-2} {\bibfield
{journal} {\bibinfo  {journal} {Nat. Commun.}\ }\textbf {\bibinfo {volume}
	{14}} (\bibinfo {year} {2023}),\ 10.1038/s41467-023-36309-2}\BibitemShut
	{NoStop}%
	\bibitem [{\citenamefont {Reyren}\ \emph {et~al.}(2007)\citenamefont {Reyren},
\citenamefont {Thiel}, \citenamefont {Caviglia}, \citenamefont {Kourkoutis},
\citenamefont {Hammerl}, \citenamefont {Richter}, \citenamefont {Schneider},
\citenamefont {Kopp}, \citenamefont {Rüetschi}, \citenamefont {Jaccard},
\citenamefont {Gabay}, \citenamefont {Muller}, \citenamefont {Triscone},\
and\ \citenamefont {Mannhart}}]{ReyrenScience2007}%
\BibitemOpen
\bibfield  {author} {\bibinfo {author} {\bibfnamefont {N.}~\bibnamefont
	{Reyren}}, \bibinfo {author} {\bibfnamefont {S.}~\bibnamefont {Thiel}},
\bibinfo {author} {\bibfnamefont {A.~D.}\ \bibnamefont {Caviglia}}, \bibinfo
{author} {\bibfnamefont {L.~Fitting}\ \bibnamefont {Kourkoutis}}, \bibinfo
{author} {\bibfnamefont {G.}~\bibnamefont {Hammerl}}, \bibinfo {author}
{\bibfnamefont {C.}~\bibnamefont {Richter}}, \bibinfo {author} {\bibfnamefont
	{C.~W.}\ \bibnamefont {Schneider}}, \bibinfo {author} {\bibfnamefont
	{T.}~\bibnamefont {Kopp}}, \bibinfo {author} {\bibfnamefont {A.-S.}\
	\bibnamefont {Rüetschi}}, \bibinfo {author} {\bibfnamefont {D.}~\bibnamefont
	{Jaccard}}, \bibinfo {author} {\bibfnamefont {M.}~\bibnamefont {Gabay}},
\bibinfo {author} {\bibfnamefont {D.~A.}\ \bibnamefont {Muller}}, \bibinfo
{author} {\bibfnamefont {J.-M.}\ \bibnamefont {Triscone}}, \ and\ \bibinfo
{author} {\bibfnamefont {J.}~\bibnamefont {Mannhart}},\ }\bibfield  {title}
{\enquote {\bibinfo {title} {Superconducting interfaces between insulating
		oxides},}\ }\href {\doibase 10.1126/science.1146006} {\bibfield  {journal}
{\bibinfo  {journal} {Science}\ }\textbf {\bibinfo {volume} {317}},\ \bibinfo
{pages} {1196--1199} (\bibinfo {year} {2007})}\BibitemShut {NoStop}%
\bibitem [{\citenamefont {Santander-Syro}\ \emph {et~al.}(2011)\citenamefont
{Santander-Syro}, \citenamefont {Copie}, \citenamefont {Kondo}, \citenamefont
{Fortuna}, \citenamefont {Pailh{\`e}s}, \citenamefont {Weht}, \citenamefont
{Qiu}, \citenamefont {Bertran}, \citenamefont {Nicolaou}, \citenamefont
{Taleb-Ibrahimi}, \citenamefont {Le~F{\`e}vre}, \citenamefont {Herranz},
\citenamefont {Bibes}, \citenamefont {Reyren}, \citenamefont {Apertet},
\citenamefont {Lecoeur}, \citenamefont {Barth{\'e}l{\'e}my},\ and\
\citenamefont {Rozenberg}}]{SantanderNature2011}%
\BibitemOpen
\bibfield  {author} {\bibinfo {author} {\bibfnamefont {A.~F.}\ \bibnamefont
	{Santander-Syro}}, \bibinfo {author} {\bibfnamefont {O.}~\bibnamefont
	{Copie}}, \bibinfo {author} {\bibfnamefont {T.}~\bibnamefont {Kondo}},
\bibinfo {author} {\bibfnamefont {F.}~\bibnamefont {Fortuna}}, \bibinfo
{author} {\bibfnamefont {S.}~\bibnamefont {Pailh{\`e}s}}, \bibinfo {author}
{\bibfnamefont {R.}~\bibnamefont {Weht}}, \bibinfo {author} {\bibfnamefont
	{X.~G.}\ \bibnamefont {Qiu}}, \bibinfo {author} {\bibfnamefont
	{F.}~\bibnamefont {Bertran}}, \bibinfo {author} {\bibfnamefont
	{A.}~\bibnamefont {Nicolaou}}, \bibinfo {author} {\bibfnamefont
	{A.}~\bibnamefont {Taleb-Ibrahimi}}, \bibinfo {author} {\bibfnamefont
	{P.}~\bibnamefont {Le~F{\`e}vre}}, \bibinfo {author} {\bibfnamefont
	{G.}~\bibnamefont {Herranz}}, \bibinfo {author} {\bibfnamefont
	{M.}~\bibnamefont {Bibes}}, \bibinfo {author} {\bibfnamefont
	{N.}~\bibnamefont {Reyren}}, \bibinfo {author} {\bibfnamefont
	{Y.}~\bibnamefont {Apertet}}, \bibinfo {author} {\bibfnamefont
	{P.}~\bibnamefont {Lecoeur}}, \bibinfo {author} {\bibfnamefont
	{A.}~\bibnamefont {Barth{\'e}l{\'e}my}}, \ and\ \bibinfo {author}
{\bibfnamefont {M.~J.}\ \bibnamefont {Rozenberg}},\ }\bibfield  {title}
{\enquote {\bibinfo {title} {{Two-dimensional electron gas with universal
			subbands at the surface of SrTiO$_3$}},}\ }\href {\doibase
10.1038/nature09720} {\bibfield  {journal} {\bibinfo  {journal} {Nature}\
}\textbf {\bibinfo {volume} {469}},\ \bibinfo {pages} {189--193} (\bibinfo
{year} {2011})}\BibitemShut {NoStop}%
\bibitem [{\citenamefont {Moser}\ \emph {et~al.}(2013)\citenamefont {Moser},
\citenamefont {Moreschini}, \citenamefont {Ja\ifmmode \acute{c}\else
	\'{c}\fi{}imovi\ifmmode~\acute{c}\else \'{c}\fi{}}, \citenamefont
{Bari\ifmmode \check{s}\else \v{s}\fi{}i\ifmmode~\acute{c}\else \'{c}\fi{}},
\citenamefont {Berger}, \citenamefont {Magrez}, \citenamefont {Chang},
\citenamefont {Kim}, \citenamefont {Bostwick}, \citenamefont {Rotenberg},
\citenamefont {Forr\'o},\ and\ \citenamefont {Grioni}}]{MoserPRL2013}%
\BibitemOpen
\bibfield  {author} {\bibinfo {author} {\bibfnamefont {S.}~\bibnamefont
	{Moser}}, \bibinfo {author} {\bibfnamefont {L.}~\bibnamefont {Moreschini}},
\bibinfo {author} {\bibfnamefont {J.}~\bibnamefont {Ja\ifmmode \acute{c}\else
		\'{c}\fi{}imovi\ifmmode~\acute{c}\else \'{c}\fi{}}}, \bibinfo {author}
{\bibfnamefont {O.~S.}\ \bibnamefont {Bari\ifmmode \check{s}\else
		\v{s}\fi{}i\ifmmode~\acute{c}\else \'{c}\fi{}}}, \bibinfo {author}
{\bibfnamefont {H.}~\bibnamefont {Berger}}, \bibinfo {author} {\bibfnamefont
	{A.}~\bibnamefont {Magrez}}, \bibinfo {author} {\bibfnamefont {Y.~J.}\
	\bibnamefont {Chang}}, \bibinfo {author} {\bibfnamefont {K.~S.}\ \bibnamefont
	{Kim}}, \bibinfo {author} {\bibfnamefont {A.}~\bibnamefont {Bostwick}},
\bibinfo {author} {\bibfnamefont {E.}~\bibnamefont {Rotenberg}}, \bibinfo
{author} {\bibfnamefont {L.}~\bibnamefont {Forr\'o}}, \ and\ \bibinfo
{author} {\bibfnamefont {M.}~\bibnamefont {Grioni}},\ }\bibfield  {title}
{\enquote {\bibinfo {title} {{Tunable Polaronic Conduction in Anatase
			${\mathrm{TiO}}_{2}$}},}\ }\href {\doibase 10.1103/PhysRevLett.110.196403}
			{\bibfield  {journal} {\bibinfo  {journal} {Phys. Rev. Lett.}\ }\textbf
{\bibinfo {volume} {110}},\ \bibinfo {pages} {196403} (\bibinfo {year}
{2013})}\BibitemShut {NoStop}%
\bibitem [{\citenamefont {Woerle}\ \emph {et~al.}(2017)\citenamefont {Woerle},
\citenamefont {Bisti}, \citenamefont {Husanu}, \citenamefont {Strocov},
\citenamefont {Schneider}, \citenamefont {Sigg}, \citenamefont {Gobrecht},
\citenamefont {Grossner},\ and\ \citenamefont {Camarda}}]{Woerle2017}%
\BibitemOpen
\bibfield  {author} {\bibinfo {author} {\bibfnamefont {J.}~\bibnamefont
	{Woerle}}, \bibinfo {author} {\bibfnamefont {F.}~\bibnamefont {Bisti}},
\bibinfo {author} {\bibfnamefont {M.-A.}\ \bibnamefont {Husanu}}, \bibinfo
{author} {\bibfnamefont {V.~N.}\ \bibnamefont {Strocov}}, \bibinfo {author}
{\bibfnamefont {C.~W.}\ \bibnamefont {Schneider}}, \bibinfo {author}
{\bibfnamefont {H.}~\bibnamefont {Sigg}}, \bibinfo {author} {\bibfnamefont
	{J.}~\bibnamefont {Gobrecht}}, \bibinfo {author} {\bibfnamefont
	{U.}~\bibnamefont {Grossner}}, \ and\ \bibinfo {author} {\bibfnamefont
	{M.}~\bibnamefont {Camarda}},\ }\bibfield  {title} {\enquote {\bibinfo
	{title} {Electronic band structure of the buried
		{SiO$_2$}/{SiC} interface investigated
		by soft x-ray {ARPES}},}\ }\href {\doibase 10.1063/1.4979102} {\bibfield
{journal} {\bibinfo  {journal} {Applied Physics Letters}\ }\textbf {\bibinfo
	{volume} {110}},\ \bibinfo {pages} {132101} (\bibinfo {year}
{2017})}\BibitemShut {NoStop}%
\bibitem [{\citenamefont {Kim}\ \emph {et~al.}(2014)\citenamefont {Kim},
\citenamefont {Krupin}, \citenamefont {Denlinger}, \citenamefont {Bostwick},
\citenamefont {Rotenberg}, \citenamefont {Zhao}, \citenamefont {Mitchell},
\citenamefont {Allen},\ and\ \citenamefont {Kim}}]{KimScience2014}%
\BibitemOpen
\bibfield  {author} {\bibinfo {author} {\bibfnamefont {Y.~K.}\ \bibnamefont
	{Kim}}, \bibinfo {author} {\bibfnamefont {O.}~\bibnamefont {Krupin}},
\bibinfo {author} {\bibfnamefont {J.~D.}\ \bibnamefont {Denlinger}}, \bibinfo
{author} {\bibfnamefont {A.}~\bibnamefont {Bostwick}}, \bibinfo {author}
{\bibfnamefont {E.}~\bibnamefont {Rotenberg}}, \bibinfo {author}
{\bibfnamefont {Q.}~\bibnamefont {Zhao}}, \bibinfo {author} {\bibfnamefont
	{J.~F.}\ \bibnamefont {Mitchell}}, \bibinfo {author} {\bibfnamefont {J.~W.}\
	\bibnamefont {Allen}}, \ and\ \bibinfo {author} {\bibfnamefont {B.~J.}\
	\bibnamefont {Kim}},\ }\bibfield  {title} {\enquote {\bibinfo {title} {{Fermi
			arcs in a doped pseudospin-1/2 Heisenberg antiferromagnet}},}\ }\href
			{\doibase 10.1126/science.1251151} {\bibfield  {journal} {\bibinfo  {journal}
	{Science}\ }\textbf {\bibinfo {volume} {345}},\ \bibinfo {pages} {187--190}
(\bibinfo {year} {2014})}\BibitemShut {NoStop}%
\bibitem [{\citenamefont {Alidoust}\ \emph {et~al.}(2014)\citenamefont
{Alidoust}, \citenamefont {Bian}, \citenamefont {Xu}, \citenamefont {Sankar},
\citenamefont {Neupane}, \citenamefont {Liu}, \citenamefont {Belopolski},
\citenamefont {Qu}, \citenamefont {Denlinger}, \citenamefont {Chou},\ and\
\citenamefont {Hasan}}]{AlidoustNatCommun2014}%
\BibitemOpen
\bibfield  {author} {\bibinfo {author} {\bibfnamefont {Nasser}\ \bibnamefont
	{Alidoust}}, \bibinfo {author} {\bibfnamefont {Guang}\ \bibnamefont {Bian}},
\bibinfo {author} {\bibfnamefont {Su-Yang}\ \bibnamefont {Xu}}, \bibinfo
{author} {\bibfnamefont {Raman}\ \bibnamefont {Sankar}}, \bibinfo {author}
{\bibfnamefont {Madhab}\ \bibnamefont {Neupane}}, \bibinfo {author}
{\bibfnamefont {Chang}\ \bibnamefont {Liu}}, \bibinfo {author} {\bibfnamefont
	{Ilya}\ \bibnamefont {Belopolski}}, \bibinfo {author} {\bibfnamefont
	{Dong-Xia}\ \bibnamefont {Qu}}, \bibinfo {author} {\bibfnamefont
	{Jonathan~D.}\ \bibnamefont {Denlinger}}, \bibinfo {author} {\bibfnamefont
	{Fang-Cheng}\ \bibnamefont {Chou}}, \ and\ \bibinfo {author} {\bibfnamefont
	{M.~Zahid}\ \bibnamefont {Hasan}},\ }\bibfield  {title} {\enquote {\bibinfo
	{title} {{Observation of monolayer valence band spin-orbit effect and induced
			quantum well states in MoX$_2$}},}\ }\href {\doibase 10.1038/ncomms5673}
			{\bibfield  {journal} {\bibinfo  {journal} {Nat. Commun.}\ }\textbf {\bibinfo
	{volume} {5}},\ \bibinfo {pages} {4673} (\bibinfo {year} {2014})}\BibitemShut
	{NoStop}%
	\bibitem [{\citenamefont {Eknapakul}\ \emph {et~al.}(2016)\citenamefont
{Eknapakul}, \citenamefont {Fongkaew}, \citenamefont {Siriroj}, \citenamefont
{Vidyasagar}, \citenamefont {Denlinger}, \citenamefont {Bawden},
\citenamefont {Mo}, \citenamefont {King}, \citenamefont {Takagi},
\citenamefont {Limpijumnong},\ and\ \citenamefont
{Meevasana}}]{EknapakulPRB2016}%
\BibitemOpen
\bibfield  {author} {\bibinfo {author} {\bibfnamefont {T.}~\bibnamefont
	{Eknapakul}}, \bibinfo {author} {\bibfnamefont {I.}~\bibnamefont {Fongkaew}},
\bibinfo {author} {\bibfnamefont {S.}~\bibnamefont {Siriroj}}, \bibinfo
{author} {\bibfnamefont {R.}~\bibnamefont {Vidyasagar}}, \bibinfo {author}
{\bibfnamefont {J.~D.}\ \bibnamefont {Denlinger}}, \bibinfo {author}
{\bibfnamefont {L.}~\bibnamefont {Bawden}}, \bibinfo {author} {\bibfnamefont
	{S.-K.}\ \bibnamefont {Mo}}, \bibinfo {author} {\bibfnamefont {P.~D.~C.}\
	\bibnamefont {King}}, \bibinfo {author} {\bibfnamefont {H.}~\bibnamefont
	{Takagi}}, \bibinfo {author} {\bibfnamefont {S.}~\bibnamefont
	{Limpijumnong}}, \ and\ \bibinfo {author} {\bibfnamefont {W.}~\bibnamefont
	{Meevasana}},\ }\bibfield  {title} {\enquote {\bibinfo {title}
	{{Nearly-free-electron system of monolayer Na on the surface of
			single-crystal ${\mathrm{HfSe}}_{2}$}},}\ }\href {\doibase
10.1103/PhysRevB.94.201121} {\bibfield  {journal} {\bibinfo  {journal} {Phys.
		Rev. B}\ }\textbf {\bibinfo {volume} {94}},\ \bibinfo {pages} {201121}
(\bibinfo {year} {2016})}\BibitemShut {NoStop}%
\bibitem [{\citenamefont {Akikubo}\ \emph {et~al.}(2016)\citenamefont
{Akikubo}, \citenamefont {Matsuda}, \citenamefont {Schmaus}, \citenamefont
{Marcaud}, \citenamefont {Liu}, \citenamefont {Silly}, \citenamefont
{Sirotti},\ and\ \citenamefont {D'Angelo}}]{AkikuboTSF2016}%
\BibitemOpen
\bibfield  {author} {\bibinfo {author} {\bibfnamefont {Kazuma}\ \bibnamefont
	{Akikubo}}, \bibinfo {author} {\bibfnamefont {Iwao}\ \bibnamefont {Matsuda}},
\bibinfo {author} {\bibfnamefont {Didier}\ \bibnamefont {Schmaus}}, \bibinfo
{author} {\bibfnamefont {Guillaume}\ \bibnamefont {Marcaud}}, \bibinfo
{author} {\bibfnamefont {Ro-Ya}\ \bibnamefont {Liu}}, \bibinfo {author}
{\bibfnamefont {Mathieu~G.}\ \bibnamefont {Silly}}, \bibinfo {author}
{\bibfnamefont {Fausto}\ \bibnamefont {Sirotti}}, \ and\ \bibinfo {author}
{\bibfnamefont {Marie}\ \bibnamefont {D'Angelo}},\ }\bibfield  {title}
{\enquote {\bibinfo {title} {{Observation of an $e_g$-derived metallic band
			at the Cs/SrTiO$_3$ interface by polarization-dependent photoemission
			spectroscopy}},}\ }\href {\doibase https://doi.org/10.1016/j.tsf.2016.01.042}
			{\bibfield  {journal} {\bibinfo  {journal} {Thin Solid Films}\ }\textbf
{\bibinfo {volume} {603}},\ \bibinfo {pages} {149--153} (\bibinfo {year}
{2016})}\BibitemShut {NoStop}%
\bibitem [{\citenamefont {Yukawa}\ \emph {et~al.}(2018)\citenamefont {Yukawa},
\citenamefont {Minohara}, \citenamefont {Shiga}, \citenamefont {Kitamura},
\citenamefont {Mitsuhashi}, \citenamefont {Kobayashi}, \citenamefont
{Horiba},\ and\ \citenamefont {Kumigashira}}]{YukawaPRB2018}%
\BibitemOpen
\bibfield  {author} {\bibinfo {author} {\bibfnamefont {R.}~\bibnamefont
	{Yukawa}}, \bibinfo {author} {\bibfnamefont {M.}~\bibnamefont {Minohara}},
\bibinfo {author} {\bibfnamefont {D.}~\bibnamefont {Shiga}}, \bibinfo
{author} {\bibfnamefont {M.}~\bibnamefont {Kitamura}}, \bibinfo {author}
{\bibfnamefont {T.}~\bibnamefont {Mitsuhashi}}, \bibinfo {author}
{\bibfnamefont {M.}~\bibnamefont {Kobayashi}}, \bibinfo {author}
{\bibfnamefont {K.}~\bibnamefont {Horiba}}, \ and\ \bibinfo {author}
{\bibfnamefont {H.}~\bibnamefont {Kumigashira}},\ }\bibfield  {title}
{\enquote {\bibinfo {title} {{Control of two-dimensional electronic states at
			anatase $\mathrm{Ti}{\mathrm{O}}_{2}\phantom{\rule{0.16em}{0ex}}(001)$
			surface by K adsorption}},}\ }\href {\doibase 10.1103/PhysRevB.97.165428}
			{\bibfield  {journal} {\bibinfo  {journal} {Phys. Rev. B}\ }\textbf {\bibinfo
	{volume} {97}},\ \bibinfo {pages} {165428} (\bibinfo {year}
{2018})}\BibitemShut {NoStop}%
\bibitem [{\citenamefont {Imada}\ \emph {et~al.}(1998)\citenamefont {Imada},
\citenamefont {Fujimori},\ and\ \citenamefont {Tokura}}]{ImadaRMP1998}%
\BibitemOpen
\bibfield  {author} {\bibinfo {author} {\bibfnamefont {Masatoshi}\
	\bibnamefont {Imada}}, \bibinfo {author} {\bibfnamefont {Atsushi}\
	\bibnamefont {Fujimori}}, \ and\ \bibinfo {author} {\bibfnamefont
	{Yoshinori}\ \bibnamefont {Tokura}},\ }\bibfield  {title} {\enquote {\bibinfo
	{title} {Metal-insulator transitions},}\ }\href {\doibase
10.1103/RevModPhys.70.1039} {\bibfield  {journal} {\bibinfo  {journal} {Rev.
		Mod. Phys.}\ }\textbf {\bibinfo {volume} {70}},\ \bibinfo {pages}
{1039--1263} (\bibinfo {year} {1998})}\BibitemShut {NoStop}%
\bibitem [{\citenamefont {Das}\ \emph {et~al.}(2018)\citenamefont {Das},
\citenamefont {Forte}, \citenamefont {Fittipaldi}, \citenamefont {Fatuzzo},
\citenamefont {Granata}, \citenamefont {Ivashko}, \citenamefont {Horio},
\citenamefont {Schindler}, \citenamefont {Dantz}, \citenamefont {Tseng},
\citenamefont {McNally}, \citenamefont {R\o{}nnow}, \citenamefont {Wan},
\citenamefont {Christensen}, \citenamefont {Pelliciari}, \citenamefont
{Olalde-Velasco}, \citenamefont {Kikugawa}, \citenamefont {Neupert},
\citenamefont {Vecchione}, \citenamefont {Schmitt}, \citenamefont {Cuoco},\
and\ \citenamefont {Chang}}]{DasPRX2018}%
\BibitemOpen
\bibfield  {author} {\bibinfo {author} {\bibfnamefont {L.}~\bibnamefont
	{Das}}, \bibinfo {author} {\bibfnamefont {F.}~\bibnamefont {Forte}}, \bibinfo
{author} {\bibfnamefont {R.}~\bibnamefont {Fittipaldi}}, \bibinfo {author}
{\bibfnamefont {C.~G.}\ \bibnamefont {Fatuzzo}}, \bibinfo {author}
{\bibfnamefont {V.}~\bibnamefont {Granata}}, \bibinfo {author} {\bibfnamefont
	{O.}~\bibnamefont {Ivashko}}, \bibinfo {author} {\bibfnamefont
	{M.}~\bibnamefont {Horio}}, \bibinfo {author} {\bibfnamefont
	{F.}~\bibnamefont {Schindler}}, \bibinfo {author} {\bibfnamefont
	{M.}~\bibnamefont {Dantz}}, \bibinfo {author} {\bibfnamefont
	{Yi}~\bibnamefont {Tseng}}, \bibinfo {author} {\bibfnamefont {D.~E.}\
	\bibnamefont {McNally}}, \bibinfo {author} {\bibfnamefont {H.~M.}\
	\bibnamefont {R\o{}nnow}}, \bibinfo {author} {\bibfnamefont {W.}~\bibnamefont
	{Wan}}, \bibinfo {author} {\bibfnamefont {N.~B.}\ \bibnamefont
	{Christensen}}, \bibinfo {author} {\bibfnamefont {J.}~\bibnamefont
	{Pelliciari}}, \bibinfo {author} {\bibfnamefont {P.}~\bibnamefont
	{Olalde-Velasco}}, \bibinfo {author} {\bibfnamefont {N.}~\bibnamefont
	{Kikugawa}}, \bibinfo {author} {\bibfnamefont {T.}~\bibnamefont {Neupert}},
\bibinfo {author} {\bibfnamefont {A.}~\bibnamefont {Vecchione}}, \bibinfo
{author} {\bibfnamefont {T.}~\bibnamefont {Schmitt}}, \bibinfo {author}
{\bibfnamefont {M.}~\bibnamefont {Cuoco}}, \ and\ \bibinfo {author}
{\bibfnamefont {J.}~\bibnamefont {Chang}},\ }\bibfield  {title} {\enquote
{\bibinfo {title} {{Spin-Orbital Excitations in
			${\mathrm{Ca}}_{2}{\mathrm{RuO}}_{4}$ Revealed by Resonant Inelastic X-Ray
			Scattering}},}\ }\href {\doibase 10.1103/PhysRevX.8.011048} {\bibfield
{journal} {\bibinfo  {journal} {Phys. Rev. X}\ }\textbf {\bibinfo {volume}
	{8}},\ \bibinfo {pages} {011048} (\bibinfo {year} {2018})}\BibitemShut
	{NoStop}%
	\bibitem [{\citenamefont {Sutter}\ \emph {et~al.}(2017)\citenamefont {Sutter},
\citenamefont {Fatuzzo}, \citenamefont {Moser}, \citenamefont {Kim},
\citenamefont {Fittipaldi}, \citenamefont {Vecchione}, \citenamefont
{Granata}, \citenamefont {Sassa}, \citenamefont {Cossalter}, \citenamefont
{Gatti}, \citenamefont {Grioni}, \citenamefont {Ronnow}, \citenamefont
{Plumb}, \citenamefont {Matt}, \citenamefont {Shi}, \citenamefont {Hoesch},
\citenamefont {Kim}, \citenamefont {Chang}, \citenamefont {Jeng},
\citenamefont {Jozwiak}, \citenamefont {Bostwick}, \citenamefont {Rotenberg},
\citenamefont {Georges}, \citenamefont {Neupert},\ and\ \citenamefont
{Chang}}]{SutterNatCom2017}%
\BibitemOpen
\bibfield  {author} {\bibinfo {author} {\bibfnamefont {D.}~\bibnamefont
	{Sutter}}, \bibinfo {author} {\bibfnamefont {C.~G.}\ \bibnamefont {Fatuzzo}},
\bibinfo {author} {\bibfnamefont {S.}~\bibnamefont {Moser}}, \bibinfo
{author} {\bibfnamefont {M.}~\bibnamefont {Kim}}, \bibinfo {author}
{\bibfnamefont {R.}~\bibnamefont {Fittipaldi}}, \bibinfo {author}
{\bibfnamefont {A.}~\bibnamefont {Vecchione}}, \bibinfo {author}
{\bibfnamefont {V.}~\bibnamefont {Granata}}, \bibinfo {author} {\bibfnamefont
	{Y.}~\bibnamefont {Sassa}}, \bibinfo {author} {\bibfnamefont
	{F.}~\bibnamefont {Cossalter}}, \bibinfo {author} {\bibfnamefont
	{G.}~\bibnamefont {Gatti}}, \bibinfo {author} {\bibfnamefont
	{M.}~\bibnamefont {Grioni}}, \bibinfo {author} {\bibfnamefont {H.~M.}\
	\bibnamefont {Ronnow}}, \bibinfo {author} {\bibfnamefont {N.~C.}\
	\bibnamefont {Plumb}}, \bibinfo {author} {\bibfnamefont {C.~E.}\ \bibnamefont
	{Matt}}, \bibinfo {author} {\bibfnamefont {M.}~\bibnamefont {Shi}}, \bibinfo
{author} {\bibfnamefont {M.}~\bibnamefont {Hoesch}}, \bibinfo {author}
{\bibfnamefont {T.~K.}\ \bibnamefont {Kim}}, \bibinfo {author} {\bibfnamefont
	{T-R.}\ \bibnamefont {Chang}}, \bibinfo {author} {\bibfnamefont {H-T.}\
	\bibnamefont {Jeng}}, \bibinfo {author} {\bibfnamefont {C.}~\bibnamefont
	{Jozwiak}}, \bibinfo {author} {\bibfnamefont {A.}~\bibnamefont {Bostwick}},
\bibinfo {author} {\bibfnamefont {E.}~\bibnamefont {Rotenberg}}, \bibinfo
{author} {\bibfnamefont {A.}~\bibnamefont {Georges}}, \bibinfo {author}
{\bibfnamefont {T.}~\bibnamefont {Neupert}}, \ and\ \bibinfo {author}
{\bibfnamefont {J.}~\bibnamefont {Chang}},\ }\bibfield  {title} {\enquote
{\bibinfo {title} {{Hallmarks of Hunds coupling in the Mott insulator
			Ca$_2$RuO$_4$}},}\ }\href {http://dx.doi.org/10.1038/ncomms15176} {\bibfield
{journal} {\bibinfo  {journal} {Nat. Commun.}\ }\textbf {\bibinfo {volume}
	{8}},\ \bibinfo {pages} {15176} (\bibinfo {year} {2017})}\BibitemShut
	{NoStop}%
	\bibitem [{\citenamefont {Ricc\`{o}}\ \emph {et~al.}(2018)\citenamefont
{Ricc\`{o}}, \citenamefont {Kim}, \citenamefont {Tamai}, \citenamefont
{Walker}, \citenamefont {Bruno}, \citenamefont {Cucchi}, \citenamefont
{Cappelli}, \citenamefont {Besnard}, \citenamefont {Kim}, \citenamefont
{Dudin}, \citenamefont {Hoesch}, \citenamefont {Gutmann}, \citenamefont
{Georges}, \citenamefont {Perry},\ and\ \citenamefont
{Baumberger}}]{RiccoNatCommun18}%
\BibitemOpen
\bibfield  {author} {\bibinfo {author} {\bibfnamefont {S.}~\bibnamefont
	{Ricc\`{o}}}, \bibinfo {author} {\bibfnamefont {M.}~\bibnamefont {Kim}},
\bibinfo {author} {\bibfnamefont {A.}~\bibnamefont {Tamai}}, \bibinfo
{author} {\bibfnamefont {S.~McKeown}\ \bibnamefont {Walker}}, \bibinfo
{author} {\bibfnamefont {F.~Y.}\ \bibnamefont {Bruno}}, \bibinfo {author}
{\bibfnamefont {I.}~\bibnamefont {Cucchi}}, \bibinfo {author} {\bibfnamefont
	{E.}~\bibnamefont {Cappelli}}, \bibinfo {author} {\bibfnamefont
	{C.}~\bibnamefont {Besnard}}, \bibinfo {author} {\bibfnamefont {T.~K.}\
	\bibnamefont {Kim}}, \bibinfo {author} {\bibfnamefont {P.}~\bibnamefont
	{Dudin}}, \bibinfo {author} {\bibfnamefont {M.}~\bibnamefont {Hoesch}},
\bibinfo {author} {\bibfnamefont {M.~J.}\ \bibnamefont {Gutmann}}, \bibinfo
{author} {\bibfnamefont {A.}~\bibnamefont {Georges}}, \bibinfo {author}
{\bibfnamefont {R.~S.}\ \bibnamefont {Perry}}, \ and\ \bibinfo {author}
{\bibfnamefont {F.}~\bibnamefont {Baumberger}},\ }\bibfield  {title}
{\enquote {\bibinfo {title} {{In situ strain tuning of the
			metal-insulator-transition of Ca$_2$RuO$_4$ in angle-resolved photoemission
			experiments}},}\ }\href {\doibase 10.1038/s41467-018-06945-0} {\bibfield
{journal} {\bibinfo  {journal} {Nat. Commun.}\ }\textbf {\bibinfo {volume}
	{9}},\ \bibinfo {pages} {4535} (\bibinfo {year} {2018})}\BibitemShut
	{NoStop}%
	\bibitem [{\citenamefont {Hu}\ \emph {et~al.}(2021)\citenamefont {Hu},
\citenamefont {Zhao}, \citenamefont {Gao}, \citenamefont {Yan}, \citenamefont
{Rong}, \citenamefont {Huang}, \citenamefont {Liu}, \citenamefont {Cai},
\citenamefont {Li}, \citenamefont {Chen}, \citenamefont {Zhao}, \citenamefont
{Liu}, \citenamefont {Jin}, \citenamefont {Xu}, \citenamefont {Xiang},\ and\
\citenamefont {Zhou}}]{HuNatCommun2021}%
\BibitemOpen
\bibfield  {author} {\bibinfo {author} {\bibfnamefont {Cheng}\ \bibnamefont
	{Hu}}, \bibinfo {author} {\bibfnamefont {Jianfa}\ \bibnamefont {Zhao}},
\bibinfo {author} {\bibfnamefont {Qiang}\ \bibnamefont {Gao}}, \bibinfo
{author} {\bibfnamefont {Hongtao}\ \bibnamefont {Yan}}, \bibinfo {author}
{\bibfnamefont {Hongtao}\ \bibnamefont {Rong}}, \bibinfo {author}
{\bibfnamefont {Jianwei}\ \bibnamefont {Huang}}, \bibinfo {author}
{\bibfnamefont {Jing}\ \bibnamefont {Liu}}, \bibinfo {author} {\bibfnamefont
	{Yongqing}\ \bibnamefont {Cai}}, \bibinfo {author} {\bibfnamefont {Cong}\
	\bibnamefont {Li}}, \bibinfo {author} {\bibfnamefont {Hao}\ \bibnamefont
	{Chen}}, \bibinfo {author} {\bibfnamefont {Lin}\ \bibnamefont {Zhao}},
\bibinfo {author} {\bibfnamefont {Guodong}\ \bibnamefont {Liu}}, \bibinfo
{author} {\bibfnamefont {Changqing}\ \bibnamefont {Jin}}, \bibinfo {author}
{\bibfnamefont {Zuyan}\ \bibnamefont {Xu}}, \bibinfo {author} {\bibfnamefont
	{Tao}\ \bibnamefont {Xiang}}, \ and\ \bibinfo {author} {\bibfnamefont
	{X.~J.}\ \bibnamefont {Zhou}},\ }\bibfield  {title} {\enquote {\bibinfo
	{title} {{Momentum-resolved visualization of electronic evolution in doping a
			Mott insulator}},}\ }\href {\doibase 10.1038/s41467-021-21605-6} {\bibfield
{journal} {\bibinfo  {journal} {Nat. Commun.}\ }\textbf {\bibinfo {volume}
	{12}},\ \bibinfo {pages} {1356} (\bibinfo {year} {2021})}\BibitemShut
	{NoStop}%
	\bibitem [{\citenamefont {Shen}\ \emph {et~al.}(2004)\citenamefont {Shen},
\citenamefont {Ronning}, \citenamefont {Lu}, \citenamefont {Lee},
\citenamefont {Ingle}, \citenamefont {Meevasana}, \citenamefont {Baumberger},
\citenamefont {Damascelli}, \citenamefont {Armitage}, \citenamefont {Miller},
\citenamefont {Kohsaka}, \citenamefont {Azuma}, \citenamefont {Takano},
\citenamefont {Takagi},\ and\ \citenamefont {Shen}}]{ShenPRL2004}%
\BibitemOpen
\bibfield  {author} {\bibinfo {author} {\bibfnamefont {K.~M.}\ \bibnamefont
	{Shen}}, \bibinfo {author} {\bibfnamefont {F.}~\bibnamefont {Ronning}},
\bibinfo {author} {\bibfnamefont {D.~H.}\ \bibnamefont {Lu}}, \bibinfo
{author} {\bibfnamefont {W.~S.}\ \bibnamefont {Lee}}, \bibinfo {author}
{\bibfnamefont {N.~J.~C.}\ \bibnamefont {Ingle}}, \bibinfo {author}
{\bibfnamefont {W.}~\bibnamefont {Meevasana}}, \bibinfo {author}
{\bibfnamefont {F.}~\bibnamefont {Baumberger}}, \bibinfo {author}
{\bibfnamefont {A.}~\bibnamefont {Damascelli}}, \bibinfo {author}
{\bibfnamefont {N.~P.}\ \bibnamefont {Armitage}}, \bibinfo {author}
{\bibfnamefont {L.~L.}\ \bibnamefont {Miller}}, \bibinfo {author}
{\bibfnamefont {Y.}~\bibnamefont {Kohsaka}}, \bibinfo {author} {\bibfnamefont
	{M.}~\bibnamefont {Azuma}}, \bibinfo {author} {\bibfnamefont
	{M.}~\bibnamefont {Takano}}, \bibinfo {author} {\bibfnamefont
	{H.}~\bibnamefont {Takagi}}, \ and\ \bibinfo {author} {\bibfnamefont {Z.-X.}\
	\bibnamefont {Shen}},\ }\bibfield  {title} {\enquote {\bibinfo {title}
	{Missing quasiparticles and the chemical potential puzzle in the doping
		evolution of the cuprate superconductors},}\ }\href {\doibase
10.1103/PhysRevLett.93.267002} {\bibfield  {journal} {\bibinfo  {journal}
	{Phys. Rev. Lett.}\ }\textbf {\bibinfo {volume} {93}},\ \bibinfo {pages}
{267002} (\bibinfo {year} {2004})}\BibitemShut {NoStop}%
\bibitem [{\citenamefont {Kim}\ \emph {et~al.}(2004)\citenamefont {Kim},
\citenamefont {Noh}, \citenamefont {Kim},\ and\ \citenamefont
{Oh}}]{KimPRL2004}%
\BibitemOpen
\bibfield  {author} {\bibinfo {author} {\bibfnamefont {Hyeong-Do}\
	\bibnamefont {Kim}}, \bibinfo {author} {\bibfnamefont {Han-Jin}\ \bibnamefont
	{Noh}}, \bibinfo {author} {\bibfnamefont {K.~H.}\ \bibnamefont {Kim}}, \ and\
\bibinfo {author} {\bibfnamefont {S.-J.}\ \bibnamefont {Oh}},\ }\bibfield
{title} {\enquote {\bibinfo {title} {{Core-Level X-Ray Photoemission
			Satellites in Ruthenates: A New Mechanism Revealing The Mott Transition}},}\
			}\href {\doibase 10.1103/PhysRevLett.93.126404} {\bibfield  {journal}
{\bibinfo  {journal} {Phys. Rev. Lett.}\ }\textbf {\bibinfo {volume} {93}},\
\bibinfo {pages} {126404} (\bibinfo {year} {2004})}\BibitemShut {NoStop}%
\bibitem [{\citenamefont {Damascelli}\ \emph {et~al.}(2000)\citenamefont
{Damascelli} \emph {et~al.}}]{DamascelliPRL00}%
\BibitemOpen
\bibfield  {author} {\bibinfo {author} {\bibfnamefont {A.}~\bibnamefont
	{Damascelli}} \emph {et~al.},\ }\bibfield  {title} {\enquote {\bibinfo
	{title} {{Fermi Surface, Surface States, and Surface Reconstruction in
			${\mathrm{Sr}}_{2}{\mathrm{RuO}}_{4}$}},}\ }\href {\doibase
10.1103/PhysRevLett.85.5194} {\bibfield  {journal} {\bibinfo  {journal}
	{Phys. Rev. Lett.}\ }\textbf {\bibinfo {volume} {85}},\ \bibinfo {pages}
{5194--5197} (\bibinfo {year} {2000})}\BibitemShut {NoStop}%
\bibitem [{\citenamefont {Tamai}\ \emph {et~al.}(2008)\citenamefont {Tamai},
\citenamefont {Allan}, \citenamefont {Mercure}, \citenamefont {Meevasana},
\citenamefont {Dunkel}, \citenamefont {Lu}, \citenamefont {Perry},
\citenamefont {Mackenzie}, \citenamefont {Singh}, \citenamefont {Shen},\ and\
\citenamefont {Baumberger}}]{TamaiPRL2008}%
\BibitemOpen
\bibfield  {author} {\bibinfo {author} {\bibfnamefont {A.}~\bibnamefont
	{Tamai}}, \bibinfo {author} {\bibfnamefont {M.~P.}\ \bibnamefont {Allan}},
\bibinfo {author} {\bibfnamefont {J.~F.}\ \bibnamefont {Mercure}}, \bibinfo
{author} {\bibfnamefont {W.}~\bibnamefont {Meevasana}}, \bibinfo {author}
{\bibfnamefont {R.}~\bibnamefont {Dunkel}}, \bibinfo {author} {\bibfnamefont
	{D.~H.}\ \bibnamefont {Lu}}, \bibinfo {author} {\bibfnamefont {R.~S.}\
	\bibnamefont {Perry}}, \bibinfo {author} {\bibfnamefont {A.~P.}\ \bibnamefont
	{Mackenzie}}, \bibinfo {author} {\bibfnamefont {D.~J.}\ \bibnamefont
	{Singh}}, \bibinfo {author} {\bibfnamefont {Z.-X.}\ \bibnamefont {Shen}}, \
and\ \bibinfo {author} {\bibfnamefont {F.}~\bibnamefont {Baumberger}},\
}\bibfield  {title} {\enquote {\bibinfo {title} {{Fermi surface and van Hove
			singularities in the itinerant metamagnet
			${\mathrm{Sr}}_{3}{\mathrm{Ru}}_{2}{\mathrm{O}}_{7}$}},}\ }\href {\doibase
10.1103/PhysRevLett.101.026407} {\bibfield  {journal} {\bibinfo  {journal}
	{Phys. Rev. Lett.}\ }\textbf {\bibinfo {volume} {101}},\ \bibinfo {pages}
{026407} (\bibinfo {year} {2008})}\BibitemShut {NoStop}%
\bibitem [{\citenamefont {Liu}\ \emph {et~al.}(2018)\citenamefont {Liu},
\citenamefont {Nair}, \citenamefont {Ruf}, \citenamefont {Schlom},\ and\
\citenamefont {Shen}}]{LiuPRB2018}%
\BibitemOpen
\bibfield  {author} {\bibinfo {author} {\bibfnamefont {Yang}\ \bibnamefont
	{Liu}}, \bibinfo {author} {\bibfnamefont {Hari~P.}\ \bibnamefont {Nair}},
\bibinfo {author} {\bibfnamefont {Jacob~P.}\ \bibnamefont {Ruf}}, \bibinfo
{author} {\bibfnamefont {Darrell~G.}\ \bibnamefont {Schlom}}, \ and\ \bibinfo
{author} {\bibfnamefont {Kyle~M.}\ \bibnamefont {Shen}},\ }\bibfield  {title}
{\enquote {\bibinfo {title} {{Revealing the hidden heavy Fermi liquid in
			${\mathrm{CaRuO}}_{3}$}},}\ }\href {\doibase 10.1103/PhysRevB.98.041110}
			{\bibfield  {journal} {\bibinfo  {journal} {Phys. Rev. B}\ }\textbf {\bibinfo
	{volume} {98}},\ \bibinfo {pages} {041110} (\bibinfo {year}
{2018})}\BibitemShut {NoStop}%
\bibitem [{\citenamefont {Sutter}\ \emph {et~al.}(2019)\citenamefont {Sutter},
\citenamefont {Kim}, \citenamefont {Matt}, \citenamefont {Horio},
\citenamefont {Fittipaldi}, \citenamefont {Vecchione}, \citenamefont
{Granata}, \citenamefont {Hauser}, \citenamefont {Sassa}, \citenamefont
{Gatti}, \citenamefont {Grioni}, \citenamefont {Hoesch}, \citenamefont {Kim},
\citenamefont {Rienks}, \citenamefont {Plumb}, \citenamefont {Shi},
\citenamefont {Neupert}, \citenamefont {Georges},\ and\ \citenamefont
{Chang}}]{SutterPRB19}%
\BibitemOpen
\bibfield  {author} {\bibinfo {author} {\bibfnamefont {D.}~\bibnamefont
	{Sutter}}, \bibinfo {author} {\bibfnamefont {M.}~\bibnamefont {Kim}},
\bibinfo {author} {\bibfnamefont {C.~E.}\ \bibnamefont {Matt}}, \bibinfo
{author} {\bibfnamefont {M.}~\bibnamefont {Horio}}, \bibinfo {author}
{\bibfnamefont {R.}~\bibnamefont {Fittipaldi}}, \bibinfo {author}
{\bibfnamefont {A.}~\bibnamefont {Vecchione}}, \bibinfo {author}
{\bibfnamefont {V.}~\bibnamefont {Granata}}, \bibinfo {author} {\bibfnamefont
	{K.}~\bibnamefont {Hauser}}, \bibinfo {author} {\bibfnamefont
	{Y.}~\bibnamefont {Sassa}}, \bibinfo {author} {\bibfnamefont
	{G.}~\bibnamefont {Gatti}}, \bibinfo {author} {\bibfnamefont
	{M.}~\bibnamefont {Grioni}}, \bibinfo {author} {\bibfnamefont
	{M.}~\bibnamefont {Hoesch}}, \bibinfo {author} {\bibfnamefont {T.~K.}\
	\bibnamefont {Kim}}, \bibinfo {author} {\bibfnamefont {E.}~\bibnamefont
	{Rienks}}, \bibinfo {author} {\bibfnamefont {N.~C.}\ \bibnamefont {Plumb}},
\bibinfo {author} {\bibfnamefont {M.}~\bibnamefont {Shi}}, \bibinfo {author}
{\bibfnamefont {T.}~\bibnamefont {Neupert}}, \bibinfo {author} {\bibfnamefont
	{A.}~\bibnamefont {Georges}}, \ and\ \bibinfo {author} {\bibfnamefont
	{J.}~\bibnamefont {Chang}},\ }\bibfield  {title} {\enquote {\bibinfo {title}
	{{Orbitally selective breakdown of Fermi liquid quasiparticles in
			${\mathrm{Ca}}_{1.8}{\mathrm{Sr}}_{0.2}{\mathrm{RuO}}_{4}$}},}\ }\href
			{\doibase 10.1103/PhysRevB.99.121115} {\bibfield  {journal} {\bibinfo
	{journal} {Phys. Rev. B}\ }\textbf {\bibinfo {volume} {99}},\ \bibinfo
{pages} {121115} (\bibinfo {year} {2019})}\BibitemShut {NoStop}%
\bibitem [{\citenamefont {Horio}\ \emph {et~al.}(2021)\citenamefont {Horio},
\citenamefont {Wang}, \citenamefont {Granata}, \citenamefont {Kramer},
\citenamefont {Sassa}, \citenamefont {J\"{o}hr}, \citenamefont {Sutter},
\citenamefont {Bold}, \citenamefont {Das}, \citenamefont {Xu}, \citenamefont
{Frison}, \citenamefont {Fittipaldi}, \citenamefont {Kim}, \citenamefont
{Cacho}, \citenamefont {Rault}, \citenamefont {F\`{e}vre}, \citenamefont
{Bertran}, \citenamefont {Plumb}, \citenamefont {Shi}, \citenamefont
{Vecchione}, \citenamefont {Fischer},\ and\ \citenamefont
{Chang}}]{HorionpjQM2021}%
\BibitemOpen
\bibfield  {author} {\bibinfo {author} {\bibfnamefont {M.}~\bibnamefont
	{Horio}}, \bibinfo {author} {\bibfnamefont {Q.}~\bibnamefont {Wang}},
\bibinfo {author} {\bibfnamefont {V.}~\bibnamefont {Granata}}, \bibinfo
{author} {\bibfnamefont {K.~P.}\ \bibnamefont {Kramer}}, \bibinfo {author}
{\bibfnamefont {Y.}~\bibnamefont {Sassa}}, \bibinfo {author} {\bibfnamefont
	{S.}~\bibnamefont {J\"{o}hr}}, \bibinfo {author} {\bibfnamefont
	{D.}~\bibnamefont {Sutter}}, \bibinfo {author} {\bibfnamefont
	{A.}~\bibnamefont {Bold}}, \bibinfo {author} {\bibfnamefont {L.}~\bibnamefont
	{Das}}, \bibinfo {author} {\bibfnamefont {Y.}~\bibnamefont {Xu}}, \bibinfo
{author} {\bibfnamefont {R.}~\bibnamefont {Frison}}, \bibinfo {author}
{\bibfnamefont {R.}~\bibnamefont {Fittipaldi}}, \bibinfo {author}
{\bibfnamefont {T.~K.}\ \bibnamefont {Kim}}, \bibinfo {author} {\bibfnamefont
	{C.}~\bibnamefont {Cacho}}, \bibinfo {author} {\bibfnamefont {J.~E.}\
	\bibnamefont {Rault}}, \bibinfo {author} {\bibfnamefont {P.~Le}\ \bibnamefont
	{F\`{e}vre}}, \bibinfo {author} {\bibfnamefont {F.}~\bibnamefont {Bertran}},
\bibinfo {author} {\bibfnamefont {N.~C.}\ \bibnamefont {Plumb}}, \bibinfo
{author} {\bibfnamefont {M.}~\bibnamefont {Shi}}, \bibinfo {author}
{\bibfnamefont {A.}~\bibnamefont {Vecchione}}, \bibinfo {author}
{\bibfnamefont {M.~H.}\ \bibnamefont {Fischer}}, \ and\ \bibinfo {author}
{\bibfnamefont {J.}~\bibnamefont {Chang}},\ }\bibfield  {title} {\enquote
{\bibinfo {title} {{Electronic reconstruction forming a $C_2$-symmetric Dirac
			semimetal in Ca$_3$Ru$_2$O$_7$}},}\ }\href {\doibase
10.1038/s41535-021-00328-3} {\bibfield  {journal} {\bibinfo  {journal} {npj
		Quantum Materials}\ }\textbf {\bibinfo {volume} {6}},\ \bibinfo {pages} {29}
(\bibinfo {year} {2021})}\BibitemShut {NoStop}%
\bibitem [{\citenamefont {Kyung}\ \emph {et~al.}(2021)\citenamefont {Kyung},
\citenamefont {Kim}, \citenamefont {Kim}, \citenamefont {Kim}, \citenamefont
{Kim}, \citenamefont {Jung}, \citenamefont {Kwon}, \citenamefont {Kim},
\citenamefont {Bostwick}, \citenamefont {Denlinger}, \citenamefont
{Yoshida},\ and\ \citenamefont {Kim}}]{KyungnpjQM2021}%
\BibitemOpen
\bibfield  {author} {\bibinfo {author} {\bibfnamefont {Wonshik}\ \bibnamefont
	{Kyung}}, \bibinfo {author} {\bibfnamefont {Choong~H.}\ \bibnamefont {Kim}},
\bibinfo {author} {\bibfnamefont {Yeong~Kwan}\ \bibnamefont {Kim}}, \bibinfo
{author} {\bibfnamefont {Beomyoung}\ \bibnamefont {Kim}}, \bibinfo {author}
{\bibfnamefont {Chul}\ \bibnamefont {Kim}}, \bibinfo {author} {\bibfnamefont
	{Woobin}\ \bibnamefont {Jung}}, \bibinfo {author} {\bibfnamefont {Junyoung}\
	\bibnamefont {Kwon}}, \bibinfo {author} {\bibfnamefont {Minsoo}\ \bibnamefont
	{Kim}}, \bibinfo {author} {\bibfnamefont {Aaron}\ \bibnamefont {Bostwick}},
\bibinfo {author} {\bibfnamefont {Jonathan~D.}\ \bibnamefont {Denlinger}},
\bibinfo {author} {\bibfnamefont {Yoshiyuki}\ \bibnamefont {Yoshida}}, \ and\
\bibinfo {author} {\bibfnamefont {Changyoung}\ \bibnamefont {Kim}},\
}\bibfield  {title} {\enquote {\bibinfo {title} {Electric-field-driven
		octahedral rotation in perovskite},}\ }\href {\doibase
10.1038/s41535-020-00306-1} {\bibfield  {journal} {\bibinfo  {journal} {npj
		Quantum Materials}\ }\textbf {\bibinfo {volume} {6}},\ \bibinfo {pages} {5}
(\bibinfo {year} {2021})}\BibitemShut {NoStop}%
\bibitem [{\citenamefont {Smith}\ \emph {et~al.}(1993)\citenamefont {Smith},
\citenamefont {Wertheim}, \citenamefont {Andrews},\ and\ \citenamefont
{Chen}}]{SmithSS1993}%
\BibitemOpen
\bibfield  {author} {\bibinfo {author} {\bibfnamefont {N.V.}\ \bibnamefont
	{Smith}}, \bibinfo {author} {\bibfnamefont {G.K.}\ \bibnamefont {Wertheim}},
\bibinfo {author} {\bibfnamefont {A.B.}\ \bibnamefont {Andrews}}, \ and\
\bibinfo {author} {\bibfnamefont {C.-T.}\ \bibnamefont {Chen}},\ }\bibfield
{title} {\enquote {\bibinfo {title} {Inelastic electron mean free paths in
		the alkali metals: effect of the empty d bands},}\ }\href {\doibase
https://doi.org/10.1016/0039-6028(93)90604-I} {\bibfield  {journal} {\bibinfo
	{journal} {Surf. Sci.}\ }\textbf {\bibinfo {volume} {282}},\ \bibinfo
{pages} {L359--L363} (\bibinfo {year} {1993})}\BibitemShut {NoStop}%
\bibitem [{\citenamefont {Zhou}\ \emph {et~al.}(2016)\citenamefont {Zhou},
\citenamefont {Li}, \citenamefont {Waugh}, \citenamefont {Parham},
\citenamefont {Kim}, \citenamefont {Sears}, \citenamefont {Gomes},
\citenamefont {Kee}, \citenamefont {Kim},\ and\ \citenamefont
{Dessau}}]{ZhouPRB2016}%
\BibitemOpen
\bibfield  {author} {\bibinfo {author} {\bibfnamefont {Xiaoqing}\
	\bibnamefont {Zhou}}, \bibinfo {author} {\bibfnamefont {Haoxiang}\
	\bibnamefont {Li}}, \bibinfo {author} {\bibfnamefont {J.~A.}\ \bibnamefont
	{Waugh}}, \bibinfo {author} {\bibfnamefont {S.}~\bibnamefont {Parham}},
\bibinfo {author} {\bibfnamefont {Heung-Sik}\ \bibnamefont {Kim}}, \bibinfo
{author} {\bibfnamefont {J.~A.}\ \bibnamefont {Sears}}, \bibinfo {author}
{\bibfnamefont {A.}~\bibnamefont {Gomes}}, \bibinfo {author} {\bibfnamefont
	{Hae-Young}\ \bibnamefont {Kee}}, \bibinfo {author} {\bibfnamefont
	{Young-June}\ \bibnamefont {Kim}}, \ and\ \bibinfo {author} {\bibfnamefont
	{D.~S.}\ \bibnamefont {Dessau}},\ }\bibfield  {title} {\enquote {\bibinfo
	{title} {{Angle-resolved photoemission study of the Kitaev candidate
			$\ensuremath{\alpha}\ensuremath{-}{\mathrm{RuCl}}_{3}$}},}\ }\href {\doibase
10.1103/PhysRevB.94.161106} {\bibfield  {journal} {\bibinfo  {journal} {Phys.
		Rev. B}\ }\textbf {\bibinfo {volume} {94}},\ \bibinfo {pages} {161106}
(\bibinfo {year} {2016})}\BibitemShut {NoStop}%
\bibitem [{\citenamefont {Eskes}\ \emph {et~al.}(1991)\citenamefont {Eskes},
\citenamefont {Meinders},\ and\ \citenamefont {Sawatzky}}]{EskesPRL1991}%
\BibitemOpen
\bibfield  {author} {\bibinfo {author} {\bibfnamefont {H.}~\bibnamefont
	{Eskes}}, \bibinfo {author} {\bibfnamefont {M.~B.~J.}\ \bibnamefont
	{Meinders}}, \ and\ \bibinfo {author} {\bibfnamefont {G.~A.}\ \bibnamefont
	{Sawatzky}},\ }\bibfield  {title} {\enquote {\bibinfo {title} {Anomalous
		transfer of spectral weight in doped strongly correlated systems},}\ }\href
		{\doibase 10.1103/PhysRevLett.67.1035} {\bibfield  {journal} {\bibinfo
	{journal} {Phys. Rev. Lett.}\ }\textbf {\bibinfo {volume} {67}},\ \bibinfo
{pages} {1035--1038} (\bibinfo {year} {1991})}\BibitemShut {NoStop}%
\bibitem [{\citenamefont {Taguchi}\ \emph {et~al.}(2005)\citenamefont
{Taguchi}, \citenamefont {Chainani}, \citenamefont {Horiba}, \citenamefont
{Takata}, \citenamefont {Yabashi}, \citenamefont {Tamasaku}, \citenamefont
{Nishino}, \citenamefont {Miwa}, \citenamefont {Ishikawa}, \citenamefont
{Takeuchi}, \citenamefont {Yamamoto}, \citenamefont {Matsunami},
\citenamefont {Shin}, \citenamefont {Yokoya}, \citenamefont {Ikenaga},
\citenamefont {Kobayashi}, \citenamefont {Mochiku}, \citenamefont {Hirata},
\citenamefont {Hori}, \citenamefont {Ishii}, \citenamefont {Nakamura},\ and\
\citenamefont {Suzuki}}]{TaguchiPRL2005}%
\BibitemOpen
\bibfield  {author} {\bibinfo {author} {\bibfnamefont {M.}~\bibnamefont
	{Taguchi}}, \bibinfo {author} {\bibfnamefont {A.}~\bibnamefont {Chainani}},
\bibinfo {author} {\bibfnamefont {K.}~\bibnamefont {Horiba}}, \bibinfo
{author} {\bibfnamefont {Y.}~\bibnamefont {Takata}}, \bibinfo {author}
{\bibfnamefont {M.}~\bibnamefont {Yabashi}}, \bibinfo {author} {\bibfnamefont
	{K.}~\bibnamefont {Tamasaku}}, \bibinfo {author} {\bibfnamefont
	{Y.}~\bibnamefont {Nishino}}, \bibinfo {author} {\bibfnamefont
	{D.}~\bibnamefont {Miwa}}, \bibinfo {author} {\bibfnamefont {T.}~\bibnamefont
	{Ishikawa}}, \bibinfo {author} {\bibfnamefont {T.}~\bibnamefont {Takeuchi}},
\bibinfo {author} {\bibfnamefont {K.}~\bibnamefont {Yamamoto}}, \bibinfo
{author} {\bibfnamefont {M.}~\bibnamefont {Matsunami}}, \bibinfo {author}
{\bibfnamefont {S.}~\bibnamefont {Shin}}, \bibinfo {author} {\bibfnamefont
	{T.}~\bibnamefont {Yokoya}}, \bibinfo {author} {\bibfnamefont
	{E.}~\bibnamefont {Ikenaga}}, \bibinfo {author} {\bibfnamefont
	{K.}~\bibnamefont {Kobayashi}}, \bibinfo {author} {\bibfnamefont
	{T.}~\bibnamefont {Mochiku}}, \bibinfo {author} {\bibfnamefont
	{K.}~\bibnamefont {Hirata}}, \bibinfo {author} {\bibfnamefont
	{J.}~\bibnamefont {Hori}}, \bibinfo {author} {\bibfnamefont {K.}~\bibnamefont
	{Ishii}}, \bibinfo {author} {\bibfnamefont {F.}~\bibnamefont {Nakamura}}, \
and\ \bibinfo {author} {\bibfnamefont {T.}~\bibnamefont {Suzuki}},\
}\bibfield  {title} {\enquote {\bibinfo {title} {{Evidence for Suppressed
			Screening on the Surface of High Temperature
			${\mathrm{La}}_{2\ensuremath{-}x}{\mathrm{Sr}}_{x}{\mathrm{CuO}}_{4}$ and
			${\mathrm{Nd}}_{2\ensuremath{-}x}{\mathrm{Ce}}_{x}{\mathrm{CuO}}_{4}$
			Superconductors}},}\ }\href {\doibase 10.1103/PhysRevLett.95.177002}
			{\bibfield  {journal} {\bibinfo  {journal} {Phys. Rev. Lett.}\ }\textbf
{\bibinfo {volume} {95}},\ \bibinfo {pages} {177002} (\bibinfo {year}
{2005})}\BibitemShut {NoStop}%
\bibitem [{\citenamefont {Horio}\ \emph {et~al.}(2018)\citenamefont {Horio},
\citenamefont {Krockenberger}, \citenamefont {Yamamoto}, \citenamefont
{Yokoyama}, \citenamefont {Takubo}, \citenamefont {Hirata}, \citenamefont
{Sakamoto}, \citenamefont {Koshiishi}, \citenamefont {Yasui}, \citenamefont
{Ikenaga}, \citenamefont {Shin}, \citenamefont {Yamamoto}, \citenamefont
{Wadati},\ and\ \citenamefont {Fujimori}}]{HorioPRL2018_2}%
\BibitemOpen
\bibfield  {author} {\bibinfo {author} {\bibfnamefont {M.}~\bibnamefont
	{Horio}}, \bibinfo {author} {\bibfnamefont {Y.}~\bibnamefont
	{Krockenberger}}, \bibinfo {author} {\bibfnamefont {K.}~\bibnamefont
	{Yamamoto}}, \bibinfo {author} {\bibfnamefont {Y.}~\bibnamefont {Yokoyama}},
\bibinfo {author} {\bibfnamefont {K.}~\bibnamefont {Takubo}}, \bibinfo
{author} {\bibfnamefont {Y.}~\bibnamefont {Hirata}}, \bibinfo {author}
{\bibfnamefont {S.}~\bibnamefont {Sakamoto}}, \bibinfo {author}
{\bibfnamefont {K.}~\bibnamefont {Koshiishi}}, \bibinfo {author}
{\bibfnamefont {A.}~\bibnamefont {Yasui}}, \bibinfo {author} {\bibfnamefont
	{E.}~\bibnamefont {Ikenaga}}, \bibinfo {author} {\bibfnamefont
	{S.}~\bibnamefont {Shin}}, \bibinfo {author} {\bibfnamefont {H.}~\bibnamefont
	{Yamamoto}}, \bibinfo {author} {\bibfnamefont {H.}~\bibnamefont {Wadati}}, \
and\ \bibinfo {author} {\bibfnamefont {A.}~\bibnamefont {Fujimori}},\
}\bibfield  {title} {\enquote {\bibinfo {title} {{Electronic Structure of
			Ce-Doped and -Undoped ${\mathrm{Nd}}_{2}{\mathrm{CuO}}_{4}$ Superconducting
			Thin Films Studied by Hard X-Ray Photoemission and Soft X-Ray Absorption
			Spectroscopy}},}\ }\href {\doibase 10.1103/PhysRevLett.120.257001} {\bibfield
{journal} {\bibinfo  {journal} {Phys. Rev. Lett.}\ }\textbf {\bibinfo
	{volume} {120}},\ \bibinfo {pages} {257001} (\bibinfo {year}
{2018})}\BibitemShut {NoStop}%
\bibitem [{\citenamefont {Cox}\ \emph {et~al.}(1983)\citenamefont {Cox},
\citenamefont {Egdell}, \citenamefont {Goodenough}, \citenamefont {Hamnett},\
and\ \citenamefont {Naish}}]{Cox1983}%
\BibitemOpen
\bibfield  {author} {\bibinfo {author} {\bibfnamefont {P~A}\ \bibnamefont
	{Cox}}, \bibinfo {author} {\bibfnamefont {R~G}\ \bibnamefont {Egdell}},
\bibinfo {author} {\bibfnamefont {J~B}\ \bibnamefont {Goodenough}}, \bibinfo
{author} {\bibfnamefont {A}~\bibnamefont {Hamnett}}, \ and\ \bibinfo {author}
{\bibfnamefont {C~C}\ \bibnamefont {Naish}},\ }\bibfield  {title} {\enquote
{\bibinfo {title} {The metal-to-semiconductor transition in ternary ruthenium
		({IV}) oxides: a study by electron spectroscopy},}\ }\href {\doibase
10.1088/0022-3719/16/32/014} {\bibfield  {journal} {\bibinfo  {journal}
	{Journal of Physics C: Solid State Physics}\ }\textbf {\bibinfo {volume}
	{16}},\ \bibinfo {pages} {6221--6239} (\bibinfo {year} {1983})}\BibitemShut
	{NoStop}%
	\bibitem [{\citenamefont {Xie}\ \emph {et~al.}(2007)\citenamefont {Xie},
\citenamefont {Yang}, \citenamefont {Shen}, \citenamefont {Zhao},
\citenamefont {Ou}, \citenamefont {Wei}, \citenamefont {Gu}, \citenamefont
{Arita}, \citenamefont {Qiao}, \citenamefont {Namatame}, \citenamefont
{Taniguchi}, \citenamefont {Kaneko}, \citenamefont {Eisaki}, \citenamefont
{Tsuei}, \citenamefont {Cheng}, \citenamefont {Vobornik}, \citenamefont
{Fujii}, \citenamefont {Rossi}, \citenamefont {Yang},\ and\ \citenamefont
{Feng}}]{XiePRL2007}%
\BibitemOpen
\bibfield  {author} {\bibinfo {author} {\bibfnamefont {B.~P.}\ \bibnamefont
	{Xie}}, \bibinfo {author} {\bibfnamefont {K.}~\bibnamefont {Yang}}, \bibinfo
{author} {\bibfnamefont {D.~W.}\ \bibnamefont {Shen}}, \bibinfo {author}
{\bibfnamefont {J.~F.}\ \bibnamefont {Zhao}}, \bibinfo {author}
{\bibfnamefont {H.~W.}\ \bibnamefont {Ou}}, \bibinfo {author} {\bibfnamefont
	{J.}~\bibnamefont {Wei}}, \bibinfo {author} {\bibfnamefont {S.~Y.}\
	\bibnamefont {Gu}}, \bibinfo {author} {\bibfnamefont {M.}~\bibnamefont
	{Arita}}, \bibinfo {author} {\bibfnamefont {S.}~\bibnamefont {Qiao}},
\bibinfo {author} {\bibfnamefont {H.}~\bibnamefont {Namatame}}, \bibinfo
{author} {\bibfnamefont {M.}~\bibnamefont {Taniguchi}}, \bibinfo {author}
{\bibfnamefont {N.}~\bibnamefont {Kaneko}}, \bibinfo {author} {\bibfnamefont
	{H.}~\bibnamefont {Eisaki}}, \bibinfo {author} {\bibfnamefont {K.~D.}\
	\bibnamefont {Tsuei}}, \bibinfo {author} {\bibfnamefont {C.~M.}\ \bibnamefont
	{Cheng}}, \bibinfo {author} {\bibfnamefont {I.}~\bibnamefont {Vobornik}},
\bibinfo {author} {\bibfnamefont {J.}~\bibnamefont {Fujii}}, \bibinfo
{author} {\bibfnamefont {G.}~\bibnamefont {Rossi}}, \bibinfo {author}
{\bibfnamefont {Z.~Q.}\ \bibnamefont {Yang}}, \ and\ \bibinfo {author}
{\bibfnamefont {D.~L.}\ \bibnamefont {Feng}},\ }\bibfield  {title} {\enquote
{\bibinfo {title} {High-energy scale revival and giant kink in the dispersion
		of a cuprate superconductor},}\ }\href {\doibase
10.1103/PhysRevLett.98.147001} {\bibfield  {journal} {\bibinfo  {journal}
	{Phys. Rev. Lett.}\ }\textbf {\bibinfo {volume} {98}},\ \bibinfo {pages}
{147001} (\bibinfo {year} {2007})}\BibitemShut {NoStop}%
\bibitem [{\citenamefont {Chang}\ \emph {et~al.}(2007)\citenamefont {Chang},
\citenamefont {Pailh\'es}, \citenamefont {Shi}, \citenamefont {M\aa{}nsson},
\citenamefont {Claesson}, \citenamefont {Tjernberg}, \citenamefont {Voigt},
\citenamefont {Perez}, \citenamefont {Patthey}, \citenamefont {Momono},
\citenamefont {Oda}, \citenamefont {Ido}, \citenamefont {Schnyder},
\citenamefont {Mudry},\ and\ \citenamefont {Mesot}}]{ChangPRB2007}%
\BibitemOpen
\bibfield  {author} {\bibinfo {author} {\bibfnamefont {J.}~\bibnamefont
	{Chang}}, \bibinfo {author} {\bibfnamefont {S.}~\bibnamefont {Pailh\'es}},
\bibinfo {author} {\bibfnamefont {M.}~\bibnamefont {Shi}}, \bibinfo {author}
{\bibfnamefont {M.}~\bibnamefont {M\aa{}nsson}}, \bibinfo {author}
{\bibfnamefont {T.}~\bibnamefont {Claesson}}, \bibinfo {author}
{\bibfnamefont {O.}~\bibnamefont {Tjernberg}}, \bibinfo {author}
{\bibfnamefont {J.}~\bibnamefont {Voigt}}, \bibinfo {author} {\bibfnamefont
	{V.}~\bibnamefont {Perez}}, \bibinfo {author} {\bibfnamefont
	{L.}~\bibnamefont {Patthey}}, \bibinfo {author} {\bibfnamefont
	{N.}~\bibnamefont {Momono}}, \bibinfo {author} {\bibfnamefont
	{M.}~\bibnamefont {Oda}}, \bibinfo {author} {\bibfnamefont {M.}~\bibnamefont
	{Ido}}, \bibinfo {author} {\bibfnamefont {A.}~\bibnamefont {Schnyder}},
\bibinfo {author} {\bibfnamefont {C.}~\bibnamefont {Mudry}}, \ and\ \bibinfo
{author} {\bibfnamefont {J.}~\bibnamefont {Mesot}},\ }\bibfield  {title}
{\enquote {\bibinfo {title} {When low- and high-energy electronic responses
		meet in cuprate superconductors},}\ }\href {\doibase
10.1103/PhysRevB.75.224508} {\bibfield  {journal} {\bibinfo  {journal} {Phys.
		Rev. B}\ }\textbf {\bibinfo {volume} {75}},\ \bibinfo {pages} {224508}
(\bibinfo {year} {2007})}\BibitemShut {NoStop}%
\bibitem [{\citenamefont {Iwasawa}\ \emph {et~al.}(2012)\citenamefont
{Iwasawa}, \citenamefont {Yoshida}, \citenamefont {Hase}, \citenamefont
{Shimada}, \citenamefont {Namatame}, \citenamefont {Taniguchi},\ and\
\citenamefont {Aiura}}]{IwasawaPRL2012}%
\BibitemOpen
\bibfield  {author} {\bibinfo {author} {\bibfnamefont {H.}~\bibnamefont
	{Iwasawa}}, \bibinfo {author} {\bibfnamefont {Y.}~\bibnamefont {Yoshida}},
\bibinfo {author} {\bibfnamefont {I.}~\bibnamefont {Hase}}, \bibinfo {author}
{\bibfnamefont {K.}~\bibnamefont {Shimada}}, \bibinfo {author} {\bibfnamefont
	{H.}~\bibnamefont {Namatame}}, \bibinfo {author} {\bibfnamefont
	{M.}~\bibnamefont {Taniguchi}}, \ and\ \bibinfo {author} {\bibfnamefont
	{Y.}~\bibnamefont {Aiura}},\ }\bibfield  {title} {\enquote {\bibinfo {title}
	{High-energy anomaly in the band dispersion of the ruthenate
		superconductor},}\ }\href {\doibase 10.1103/PhysRevLett.109.066404}
		{\bibfield  {journal} {\bibinfo  {journal} {Phys. Rev. Lett.}\ }\textbf
{\bibinfo {volume} {109}},\ \bibinfo {pages} {066404} (\bibinfo {year}
{2012})}\BibitemShut {NoStop}%
\bibitem [{\citenamefont {Fukazawa}\ and\ \citenamefont
{Maeno}(2001)}]{FukazawaJPSJ2000}%
\BibitemOpen
\bibfield  {author} {\bibinfo {author} {\bibfnamefont {Hideto}\ \bibnamefont
	{Fukazawa}}\ and\ \bibinfo {author} {\bibfnamefont {Yoshiteru}\ \bibnamefont
	{Maeno}},\ }\bibfield  {title} {\enquote {\bibinfo {title} {{Filling Control
			of the Mott Insulator Ca$_2$RuO$_4$}},}\ }\href {\doibase
10.1143/JPSJ.70.460} {\bibfield  {journal} {\bibinfo  {journal} {J. Phys.
		Soc. Jpn.}\ }\textbf {\bibinfo {volume} {70}},\ \bibinfo {pages} {460--467}
(\bibinfo {year} {2001})}\BibitemShut {NoStop}%
\bibitem [{\citenamefont {Cao}\ \emph {et~al.}(2000)\citenamefont {Cao},
\citenamefont {McCall}, \citenamefont {Dobrosavljevic}, \citenamefont
{Alexander}, \citenamefont {Crow},\ and\ \citenamefont
{Guertin}}]{CaoPRB2000}%
\BibitemOpen
\bibfield  {author} {\bibinfo {author} {\bibfnamefont {G.}~\bibnamefont
	{Cao}}, \bibinfo {author} {\bibfnamefont {S.}~\bibnamefont {McCall}},
\bibinfo {author} {\bibfnamefont {V.}~\bibnamefont {Dobrosavljevic}},
\bibinfo {author} {\bibfnamefont {C.~S.}\ \bibnamefont {Alexander}}, \bibinfo
{author} {\bibfnamefont {J.~E.}\ \bibnamefont {Crow}}, \ and\ \bibinfo
{author} {\bibfnamefont {R.~P.}\ \bibnamefont {Guertin}},\ }\bibfield
{title} {\enquote {\bibinfo {title} {{Ground-state instability of the Mott
			insulator ${\mathrm{Ca}}_{2}{\mathrm{RuO}}_{4}:$ Impact of slight La doping
			on the metal-insulator transition and magnetic ordering}},}\ }\href {\doibase
10.1103/PhysRevB.61.R5053} {\bibfield  {journal} {\bibinfo  {journal} {Phys.
		Rev. B}\ }\textbf {\bibinfo {volume} {61}},\ \bibinfo {pages} {R5053--R5057}
(\bibinfo {year} {2000})}\BibitemShut {NoStop}%
\bibitem [{\citenamefont {Pincini}\ \emph {et~al.}(2019)\citenamefont
{Pincini}, \citenamefont {Veiga}, \citenamefont {Dashwood}, \citenamefont
{Forte}, \citenamefont {Cuoco}, \citenamefont {Perry}, \citenamefont
{Bencok}, \citenamefont {Boothroyd},\ and\ \citenamefont
{McMorrow}}]{Pincini2019}%
\BibitemOpen
\bibfield  {author} {\bibinfo {author} {\bibfnamefont {D.}~\bibnamefont
	{Pincini}}, \bibinfo {author} {\bibfnamefont {L.~S.~I.}\ \bibnamefont
	{Veiga}}, \bibinfo {author} {\bibfnamefont {C.~D.}\ \bibnamefont {Dashwood}},
\bibinfo {author} {\bibfnamefont {F.}~\bibnamefont {Forte}}, \bibinfo
{author} {\bibfnamefont {M.}~\bibnamefont {Cuoco}}, \bibinfo {author}
{\bibfnamefont {R.~S.}\ \bibnamefont {Perry}}, \bibinfo {author}
{\bibfnamefont {P.}~\bibnamefont {Bencok}}, \bibinfo {author} {\bibfnamefont
	{A.~T.}\ \bibnamefont {Boothroyd}}, \ and\ \bibinfo {author} {\bibfnamefont
	{D.~F.}\ \bibnamefont {McMorrow}},\ }\bibfield  {title} {\enquote {\bibinfo
	{title} {{Tuning of the ${\mathrm{Ru}}^{4+}$ ground-state orbital population
			in the $4{d}^{4}$ Mott insulator ${\mathrm{Ca}}_{2}{\mathrm{RuO}}_{4}$
			achieved by La doping}},}\ }\href {\doibase 10.1103/PhysRevB.99.075125}
			{\bibfield  {journal} {\bibinfo  {journal} {Phys. Rev. B}\ }\textbf {\bibinfo
	{volume} {99}},\ \bibinfo {pages} {075125} (\bibinfo {year}
{2019})}\BibitemShut {NoStop}%
\bibitem [{\citenamefont {Nakatsuji}\ and\ \citenamefont
{Maeno}(2000)}]{NakatsujiPRL00}%
\BibitemOpen
\bibfield  {author} {\bibinfo {author} {\bibfnamefont {S.}~\bibnamefont
	{Nakatsuji}}\ and\ \bibinfo {author} {\bibfnamefont {Y.}~\bibnamefont
	{Maeno}},\ }\bibfield  {title} {\enquote {\bibinfo {title}
	{{Quasi-two-dimensional Mott transition system
			${\mathrm{Ca}}_{2\ensuremath{-}\mathit{x}}{\mathrm{Sr}}_{\mathit{x}}{\mathrm{RuO}}_{4}$}},}\
			}\href {\doibase 10.1103/PhysRevLett.84.2666} {\bibfield  {journal} {\bibinfo
	{journal} {Phys. Rev. Lett.}\ }\textbf {\bibinfo {volume} {84}},\ \bibinfo
{pages} {2666--2669} (\bibinfo {year} {2000})}\BibitemShut {NoStop}%
\bibitem [{\citenamefont {Armitage}\ \emph {et~al.}(2002)\citenamefont
{Armitage}, \citenamefont {Ronning}, \citenamefont {Lu}, \citenamefont {Kim},
\citenamefont {Damascelli}, \citenamefont {Shen}, \citenamefont {Feng},
\citenamefont {Eisaki}, \citenamefont {Shen}, \citenamefont {Mang},
\citenamefont {Kaneko}, \citenamefont {Greven}, \citenamefont {Onose},
\citenamefont {Taguchi},\ and\ \citenamefont {Tokura}}]{ArmitagePRL2002}%
\BibitemOpen
\bibfield  {author} {\bibinfo {author} {\bibfnamefont {N.~P.}\ \bibnamefont
	{Armitage}}, \bibinfo {author} {\bibfnamefont {F.}~\bibnamefont {Ronning}},
\bibinfo {author} {\bibfnamefont {D.~H.}\ \bibnamefont {Lu}}, \bibinfo
{author} {\bibfnamefont {C.}~\bibnamefont {Kim}}, \bibinfo {author}
{\bibfnamefont {A.}~\bibnamefont {Damascelli}}, \bibinfo {author}
{\bibfnamefont {K.~M.}\ \bibnamefont {Shen}}, \bibinfo {author}
{\bibfnamefont {D.~L.}\ \bibnamefont {Feng}}, \bibinfo {author}
{\bibfnamefont {H.}~\bibnamefont {Eisaki}}, \bibinfo {author} {\bibfnamefont
	{Z.-X.}\ \bibnamefont {Shen}}, \bibinfo {author} {\bibfnamefont {P.~K.}\
	\bibnamefont {Mang}}, \bibinfo {author} {\bibfnamefont {N.}~\bibnamefont
	{Kaneko}}, \bibinfo {author} {\bibfnamefont {M.}~\bibnamefont {Greven}},
\bibinfo {author} {\bibfnamefont {Y.}~\bibnamefont {Onose}}, \bibinfo
{author} {\bibfnamefont {Y.}~\bibnamefont {Taguchi}}, \ and\ \bibinfo
{author} {\bibfnamefont {Y.}~\bibnamefont {Tokura}},\ }\bibfield  {title}
{\enquote {\bibinfo {title} {Doping dependence of an $\mathit{n}$-type
		cuprate superconductor investigated by angle-resolved photoemission
		spectroscopy},}\ }\href {\doibase 10.1103/PhysRevLett.88.257001} {\bibfield
{journal} {\bibinfo  {journal} {Phys. Rev. Lett.}\ }\textbf {\bibinfo
	{volume} {88}},\ \bibinfo {pages} {257001} (\bibinfo {year}
{2002})}\BibitemShut {NoStop}%
\bibitem [{\citenamefont {Soukiassian}\ \emph {et~al.}(1985)\citenamefont
{Soukiassian}, \citenamefont {Riwan}, \citenamefont {Lecante}, \citenamefont
{Wimmer}, \citenamefont {Chubb},\ and\ \citenamefont
{Freeman}}]{SoukiassianPRB1985}%
\BibitemOpen
\bibfield  {author} {\bibinfo {author} {\bibfnamefont {P.}~\bibnamefont
	{Soukiassian}}, \bibinfo {author} {\bibfnamefont {R.}~\bibnamefont {Riwan}},
\bibinfo {author} {\bibfnamefont {J.}~\bibnamefont {Lecante}}, \bibinfo
{author} {\bibfnamefont {E.}~\bibnamefont {Wimmer}}, \bibinfo {author}
{\bibfnamefont {S.~R.}\ \bibnamefont {Chubb}}, \ and\ \bibinfo {author}
{\bibfnamefont {A.~J.}\ \bibnamefont {Freeman}},\ }\bibfield  {title}
{\enquote {\bibinfo {title} {Adsorbate-induced shifts of electronic surface
		states: Cs on the (100) faces of tungsten, molybdenum, and tantalum},}\
		}\href {\doibase 10.1103/PhysRevB.31.4911} {\bibfield  {journal} {\bibinfo
	{journal} {Phys. Rev. B}\ }\textbf {\bibinfo {volume} {31}},\ \bibinfo
{pages} {4911--4923} (\bibinfo {year} {1985})}\BibitemShut {NoStop}%
\bibitem [{\citenamefont {Ozawa}\ \emph {et~al.}(1997)\citenamefont {Ozawa},
\citenamefont {Ishikawa}, \citenamefont {Miyazaki}, \citenamefont {Edamoto},
\citenamefont {Kato},\ and\ \citenamefont {Otani}}]{OzawaSS1997}%
\BibitemOpen
\bibfield  {author} {\bibinfo {author} {\bibfnamefont {K.}~\bibnamefont
	{Ozawa}}, \bibinfo {author} {\bibfnamefont {S.}~\bibnamefont {Ishikawa}},
\bibinfo {author} {\bibfnamefont {E.}~\bibnamefont {Miyazaki}}, \bibinfo
{author} {\bibfnamefont {K.}~\bibnamefont {Edamoto}}, \bibinfo {author}
{\bibfnamefont {H.}~\bibnamefont {Kato}}, \ and\ \bibinfo {author}
{\bibfnamefont {S.}~\bibnamefont {Otani}},\ }\bibfield  {title} {\enquote
{\bibinfo {title} {{Potassium adsorption on the polar NbC(111) surface:
			angle-resolved photoemission study}},}\ }\href {\doibase
https://doi.org/10.1016/S0039-6028(97)01299-5} {\bibfield  {journal}
{\bibinfo  {journal} {Surf. Sci.}\ }\textbf {\bibinfo {volume} {375}},\
\bibinfo {pages} {250--256} (\bibinfo {year} {1997})}\BibitemShut {NoStop}%
\bibitem [{\citenamefont {Ozawa}\ \emph {et~al.}(1999)\citenamefont {Ozawa},
\citenamefont {Noda}, \citenamefont {Nakane}, \citenamefont {Yamazaki},
\citenamefont {Edamoto},\ and\ \citenamefont {Otani}}]{OzawaSS1999}%
\BibitemOpen
\bibfield  {author} {\bibinfo {author} {\bibfnamefont {K.}~\bibnamefont
	{Ozawa}}, \bibinfo {author} {\bibfnamefont {T.}~\bibnamefont {Noda}},
\bibinfo {author} {\bibfnamefont {T.}~\bibnamefont {Nakane}}, \bibinfo
{author} {\bibfnamefont {M.}~\bibnamefont {Yamazaki}}, \bibinfo {author}
{\bibfnamefont {K.}~\bibnamefont {Edamoto}}, \ and\ \bibinfo {author}
{\bibfnamefont {S.}~\bibnamefont {Otani}},\ }\bibfield  {title} {\enquote
{\bibinfo {title} {{Photoemission study of K adsorption on ZrC(111)}},}\
}\href {\doibase https://doi.org/10.1016/S0039-6028(99)00160-0} {\bibfield
{journal} {\bibinfo  {journal} {Surf. Sci.}\ }\textbf {\bibinfo {volume}
	{433-435}},\ \bibinfo {pages} {700--704} (\bibinfo {year}
{1999})}\BibitemShut {NoStop}%
\bibitem [{\citenamefont {Zhang}\ \emph {et~al.}(2016)\citenamefont {Zhang},
\citenamefont {Wang}, \citenamefont {Cui}, \citenamefont {Wang},
\citenamefont {Argondizzo}, \citenamefont {Zhao},\ and\ \citenamefont
{Petek}}]{ZhangPRB2016}%
\BibitemOpen
\bibfield  {author} {\bibinfo {author} {\bibfnamefont {Shengmin}\
	\bibnamefont {Zhang}}, \bibinfo {author} {\bibfnamefont {Cong}\ \bibnamefont
	{Wang}}, \bibinfo {author} {\bibfnamefont {Xuefeng}\ \bibnamefont {Cui}},
\bibinfo {author} {\bibfnamefont {Yanan}\ \bibnamefont {Wang}}, \bibinfo
{author} {\bibfnamefont {Adam}\ \bibnamefont {Argondizzo}}, \bibinfo {author}
{\bibfnamefont {Jin}\ \bibnamefont {Zhao}}, \ and\ \bibinfo {author}
{\bibfnamefont {Hrvoje}\ \bibnamefont {Petek}},\ }\bibfield  {title}
{\enquote {\bibinfo {title} {{Time-resolved photoemission study of the
			electronic structure and dynamics of chemisorbed alkali atoms on
			Ru(0001)}},}\ }\href {\doibase 10.1103/PhysRevB.93.045401} {\bibfield
{journal} {\bibinfo  {journal} {Phys. Rev. B}\ }\textbf {\bibinfo {volume}
	{93}},\ \bibinfo {pages} {045401} (\bibinfo {year} {2016})}\BibitemShut
	{NoStop}%
	\bibitem [{\citenamefont {S\'en\'echal}\ \emph {et~al.}(2002)\citenamefont
{S\'en\'echal}, \citenamefont {Perez},\ and\ \citenamefont
{Plouffe}}]{Senechal2002}%
\BibitemOpen
\bibfield  {author} {\bibinfo {author} {\bibfnamefont {David}\ \bibnamefont
	{S\'en\'echal}}, \bibinfo {author} {\bibfnamefont {Danny}\ \bibnamefont
	{Perez}}, \ and\ \bibinfo {author} {\bibfnamefont {Dany}\ \bibnamefont
	{Plouffe}},\ }\bibfield  {title} {\enquote {\bibinfo {title} {Cluster
		perturbation theory for hubbard models},}\ }\href {\doibase
10.1103/PhysRevB.66.075129} {\bibfield  {journal} {\bibinfo  {journal} {Phys.
		Rev. B}\ }\textbf {\bibinfo {volume} {66}},\ \bibinfo {pages} {075129}
(\bibinfo {year} {2002})}\BibitemShut {NoStop}%
\bibitem [{\citenamefont {Sekiyama}\ \emph {et~al.}(2004)\citenamefont
{Sekiyama}, \citenamefont {Fujiwara}, \citenamefont {Imada}, \citenamefont
{Suga}, \citenamefont {Eisaki}, \citenamefont {Uchida}, \citenamefont
{Takegahara}, \citenamefont {Harima}, \citenamefont {Saitoh}, \citenamefont
{Nekrasov}, \citenamefont {Keller}, \citenamefont {Kondakov}, \citenamefont
{Kozhevnikov}, \citenamefont {Pruschke}, \citenamefont {Held}, \citenamefont
{Vollhardt},\ and\ \citenamefont {Anisimov}}]{SekiyamaPRL2004}%
\BibitemOpen
\bibfield  {author} {\bibinfo {author} {\bibfnamefont {A.}~\bibnamefont
	{Sekiyama}}, \bibinfo {author} {\bibfnamefont {H.}~\bibnamefont {Fujiwara}},
\bibinfo {author} {\bibfnamefont {S.}~\bibnamefont {Imada}}, \bibinfo
{author} {\bibfnamefont {S.}~\bibnamefont {Suga}}, \bibinfo {author}
{\bibfnamefont {H.}~\bibnamefont {Eisaki}}, \bibinfo {author} {\bibfnamefont
	{S.~I.}\ \bibnamefont {Uchida}}, \bibinfo {author} {\bibfnamefont
	{K.}~\bibnamefont {Takegahara}}, \bibinfo {author} {\bibfnamefont
	{H.}~\bibnamefont {Harima}}, \bibinfo {author} {\bibfnamefont
	{Y.}~\bibnamefont {Saitoh}}, \bibinfo {author} {\bibfnamefont {I.~A.}\
	\bibnamefont {Nekrasov}}, \bibinfo {author} {\bibfnamefont {G.}~\bibnamefont
	{Keller}}, \bibinfo {author} {\bibfnamefont {D.~E.}\ \bibnamefont
	{Kondakov}}, \bibinfo {author} {\bibfnamefont {A.~V.}\ \bibnamefont
	{Kozhevnikov}}, \bibinfo {author} {\bibfnamefont {Th.}\ \bibnamefont
	{Pruschke}}, \bibinfo {author} {\bibfnamefont {K.}~\bibnamefont {Held}},
\bibinfo {author} {\bibfnamefont {D.}~\bibnamefont {Vollhardt}}, \ and\
\bibinfo {author} {\bibfnamefont {V.~I.}\ \bibnamefont {Anisimov}},\
}\bibfield  {title} {\enquote {\bibinfo {title} {{Mutual Experimental and
			Theoretical Validation of Bulk Photoemission Spectra of
			${\mathrm{S}\mathrm{r}}_{1\ensuremath{-}x}{\mathrm{C}\mathrm{a}}_{x}{\mathrm{V}\mathrm{O}}_{3}$}},}\
			}\href {\doibase 10.1103/PhysRevLett.93.156402} {\bibfield  {journal}
{\bibinfo  {journal} {Phys. Rev. Lett.}\ }\textbf {\bibinfo {volume} {93}},\
\bibinfo {pages} {156402} (\bibinfo {year} {2004})}\BibitemShut {NoStop}%
\bibitem [{\citenamefont {Rodolakis}\ \emph {et~al.}(2009)\citenamefont
{Rodolakis}, \citenamefont {Mansart}, \citenamefont {Papalazarou},
\citenamefont {Gorovikov}, \citenamefont {Vilmercati}, \citenamefont
{Petaccia}, \citenamefont {Goldoni}, \citenamefont {Rueff}, \citenamefont
{Lupi}, \citenamefont {Metcalf},\ and\ \citenamefont
{Marsi}}]{RodolakisPRL2009}%
\BibitemOpen
\bibfield  {author} {\bibinfo {author} {\bibfnamefont {F.}~\bibnamefont
	{Rodolakis}}, \bibinfo {author} {\bibfnamefont {B.}~\bibnamefont {Mansart}},
\bibinfo {author} {\bibfnamefont {E.}~\bibnamefont {Papalazarou}}, \bibinfo
{author} {\bibfnamefont {S.}~\bibnamefont {Gorovikov}}, \bibinfo {author}
{\bibfnamefont {P.}~\bibnamefont {Vilmercati}}, \bibinfo {author}
{\bibfnamefont {L.}~\bibnamefont {Petaccia}}, \bibinfo {author}
{\bibfnamefont {A.}~\bibnamefont {Goldoni}}, \bibinfo {author} {\bibfnamefont
	{J.~P.}\ \bibnamefont {Rueff}}, \bibinfo {author} {\bibfnamefont
	{S.}~\bibnamefont {Lupi}}, \bibinfo {author} {\bibfnamefont {P.}~\bibnamefont
	{Metcalf}}, \ and\ \bibinfo {author} {\bibfnamefont {M.}~\bibnamefont
	{Marsi}},\ }\bibfield  {title} {\enquote {\bibinfo {title} {{Quasiparticles
			at the Mott Transition in ${\mathrm{V}}_{2}{\mathrm{O}}_{3}$: Wave Vector
			Dependence and Surface Attenuation}},}\ }\href {\doibase
10.1103/PhysRevLett.102.066805} {\bibfield  {journal} {\bibinfo  {journal}
	{Phys. Rev. Lett.}\ }\textbf {\bibinfo {volume} {102}},\ \bibinfo {pages}
{066805} (\bibinfo {year} {2009})}\BibitemShut {NoStop}%
\bibitem [{\citenamefont {Yoshimatsu}\ \emph {et~al.}(2010)\citenamefont
{Yoshimatsu}, \citenamefont {Okabe}, \citenamefont {Kumigashira},
\citenamefont {Okamoto}, \citenamefont {Aizaki}, \citenamefont {Fujimori},\
and\ \citenamefont {Oshima}}]{YoshimatsuPRL2010}%
\BibitemOpen
\bibfield  {author} {\bibinfo {author} {\bibfnamefont {K.}~\bibnamefont
	{Yoshimatsu}}, \bibinfo {author} {\bibfnamefont {T.}~\bibnamefont {Okabe}},
\bibinfo {author} {\bibfnamefont {H.}~\bibnamefont {Kumigashira}}, \bibinfo
{author} {\bibfnamefont {S.}~\bibnamefont {Okamoto}}, \bibinfo {author}
{\bibfnamefont {S.}~\bibnamefont {Aizaki}}, \bibinfo {author} {\bibfnamefont
	{A.}~\bibnamefont {Fujimori}}, \ and\ \bibinfo {author} {\bibfnamefont
	{M.}~\bibnamefont {Oshima}},\ }\bibfield  {title} {\enquote {\bibinfo {title}
	{{Dimensional-Crossover-Driven Metal-Insulator Transition in
			${\mathrm{SrVO}}_{3}$ Ultrathin Films}},}\ }\href {\doibase
10.1103/PhysRevLett.104.147601} {\bibfield  {journal} {\bibinfo  {journal}
	{Phys. Rev. Lett.}\ }\textbf {\bibinfo {volume} {104}},\ \bibinfo {pages}
{147601} (\bibinfo {year} {2010})}\BibitemShut {NoStop}%
\bibitem [{\citenamefont {Takizawa}\ \emph {et~al.}(2005)\citenamefont
{Takizawa}, \citenamefont {Toyota}, \citenamefont {Wadati}, \citenamefont
{Chikamatsu}, \citenamefont {Kumigashira}, \citenamefont {Fujimori},
\citenamefont {Oshima}, \citenamefont {Fang}, \citenamefont {Lippmaa},
\citenamefont {Kawasaki},\ and\ \citenamefont {Koinuma}}]{TakizawaPRB2005}%
\BibitemOpen
\bibfield  {author} {\bibinfo {author} {\bibfnamefont {M.}~\bibnamefont
	{Takizawa}}, \bibinfo {author} {\bibfnamefont {D.}~\bibnamefont {Toyota}},
\bibinfo {author} {\bibfnamefont {H.}~\bibnamefont {Wadati}}, \bibinfo
{author} {\bibfnamefont {A.}~\bibnamefont {Chikamatsu}}, \bibinfo {author}
{\bibfnamefont {H.}~\bibnamefont {Kumigashira}}, \bibinfo {author}
{\bibfnamefont {A.}~\bibnamefont {Fujimori}}, \bibinfo {author}
{\bibfnamefont {M.}~\bibnamefont {Oshima}}, \bibinfo {author} {\bibfnamefont
	{Z.}~\bibnamefont {Fang}}, \bibinfo {author} {\bibfnamefont {M.}~\bibnamefont
	{Lippmaa}}, \bibinfo {author} {\bibfnamefont {M.}~\bibnamefont {Kawasaki}}, \
and\ \bibinfo {author} {\bibfnamefont {H.}~\bibnamefont {Koinuma}},\
}\bibfield  {title} {\enquote {\bibinfo {title} {{Manifestation of
			correlation effects in the photoemission spectra of
			${\mathrm{Ca}}_{1\ensuremath{-}x}{\mathrm{Sr}}_{x}\mathrm{Ru}{\mathrm{O}}_{3}$}},}\
			}\href {\doibase 10.1103/PhysRevB.72.060404} {\bibfield  {journal} {\bibinfo
	{journal} {Phys. Rev. B}\ }\textbf {\bibinfo {volume} {72}},\ \bibinfo
{pages} {060404} (\bibinfo {year} {2005})}\BibitemShut {NoStop}%
\bibitem [{\citenamefont {Panaccione}\ \emph {et~al.}(2011)\citenamefont
{Panaccione}, \citenamefont {Manju}, \citenamefont {Offi}, \citenamefont
{Annese}, \citenamefont {Vobornik}, \citenamefont {Torelli}, \citenamefont
{Zhu}, \citenamefont {Hossain}, \citenamefont {Simonelli}, \citenamefont
{Fondacaro}, \citenamefont {Lacovig}, \citenamefont {Guarino}, \citenamefont
{Yoshida}, \citenamefont {Sawatzky},\ and\ \citenamefont
{Damascelli}}]{PanaccioneNJP2011}%
\BibitemOpen
\bibfield  {author} {\bibinfo {author} {\bibfnamefont {G}~\bibnamefont
	{Panaccione}}, \bibinfo {author} {\bibfnamefont {U}~\bibnamefont {Manju}},
\bibinfo {author} {\bibfnamefont {F}~\bibnamefont {Offi}}, \bibinfo {author}
{\bibfnamefont {E}~\bibnamefont {Annese}}, \bibinfo {author} {\bibfnamefont
	{I}~\bibnamefont {Vobornik}}, \bibinfo {author} {\bibfnamefont
	{P}~\bibnamefont {Torelli}}, \bibinfo {author} {\bibfnamefont {Z~H}\
	\bibnamefont {Zhu}}, \bibinfo {author} {\bibfnamefont {M~A}\ \bibnamefont
	{Hossain}}, \bibinfo {author} {\bibfnamefont {L}~\bibnamefont {Simonelli}},
\bibinfo {author} {\bibfnamefont {A}~\bibnamefont {Fondacaro}}, \bibinfo
{author} {\bibfnamefont {P}~\bibnamefont {Lacovig}}, \bibinfo {author}
{\bibfnamefont {A}~\bibnamefont {Guarino}}, \bibinfo {author} {\bibfnamefont
	{Y}~\bibnamefont {Yoshida}}, \bibinfo {author} {\bibfnamefont {G~A}\
	\bibnamefont {Sawatzky}}, \ and\ \bibinfo {author} {\bibfnamefont
	{A}~\bibnamefont {Damascelli}},\ }\bibfield  {title} {\enquote {\bibinfo
	{title} {{Depth dependence of itinerant character in Mn-substituted
			Sr$_3$Ru$_2$O$_7$}},}\ }\href {\doibase 10.1088/1367-2630/13/5/053059}
			{\bibfield  {journal} {\bibinfo  {journal} {New J. Phys.}\ }\textbf {\bibinfo
	{volume} {13}},\ \bibinfo {pages} {053059} (\bibinfo {year}
{2011})}\BibitemShut {NoStop}%
\bibitem [{\citenamefont {{Fukazawa}}\ \emph {et~al.}(2000)\citenamefont
{{Fukazawa}}, \citenamefont {{Nakatsuji}},\ and\ \citenamefont
{{Maeno}}}]{FukazawaPhysB00}%
\BibitemOpen
\bibfield  {author} {\bibinfo {author} {\bibfnamefont {H.}~\bibnamefont
	{{Fukazawa}}}, \bibinfo {author} {\bibfnamefont {S.}~\bibnamefont
	{{Nakatsuji}}}, \ and\ \bibinfo {author} {\bibfnamefont {Y.}~\bibnamefont
	{{Maeno}}},\ }\bibfield  {title} {\enquote {\bibinfo {title} {{Intrinsic
			Properties of the Mott Insulator Ca$_{2}$RuO$_{4+\delta}$ ($\delta=0$)
			Studied with Single Crystals}},}\ }\href {\doibase
10.1016/S0921-4526(99)00989-8} {\bibfield  {journal} {\bibinfo  {journal}
	{Physica B Condens Matter}\ }\textbf {\bibinfo {volume} {281}},\ \bibinfo
{pages} {613--614} (\bibinfo {year} {2000})}\BibitemShut {NoStop}%
\bibitem [{\citenamefont {Nakatsuji}\ and\ \citenamefont
{Maeno}(2001)}]{snakatsujiJSSCHEM2001}%
\BibitemOpen
\bibfield  {author} {\bibinfo {author} {\bibfnamefont {Satoru}\ \bibnamefont
	{Nakatsuji}}\ and\ \bibinfo {author} {\bibfnamefont {Yoshiteru}\ \bibnamefont
	{Maeno}},\ }\bibfield  {title} {\enquote {\bibinfo {title} {{Synthesis and
			Single-Crystal Growth of Ca$_{2-x}$Sr$_x$RuO$_4$}},}\ }\href
			{https://doi.org/10.1006/jssc.2000.8953} {\bibfield  {journal} {\bibinfo
	{journal} {J Solid State Chem}\ }\textbf {\bibinfo {volume} {156}},\ \bibinfo
{pages} {26 -- 31} (\bibinfo {year} {2001})}\BibitemShut {NoStop}%
\end{thebibliography}

%

\end{document}